\documentclass[aps,pra,twocolumn,groupedaddress]{revtex4-1}

\usepackage[utf8]{inputenc}
\usepackage{graphicx}
\usepackage{color}
\usepackage{amsmath}
\usepackage{bm}
\usepackage{hyperref}
\usepackage[caption=false]{subfig}
\usepackage{siunitx}

\newcommand{\beq}{\begin{equation}}
\newcommand{\eeq}{\end{equation}}

\begin{document}

\title{Stability and dynamics of ion rings in linear multipole traps}

\newcommand{\vecf}[1]{\bm{#1}}

\author{Florian Cartarius}
%\email[]{}
\affiliation{Theoretische Physik, Universit\"at des Saarlandes, 66123 Saarbr\"ucken, Germany}

\author{Cecilia Cormick}
\affiliation{Theoretische Physik, Universit\"at des Saarlandes, 66123 Saarbr\"ucken, Germany}

\author{Giovanna Morigi}
\affiliation{Theoretische Physik, Universit\"at des Saarlandes, 66123 Saarbr\"ucken, Germany}

\date{\today}

\begin{abstract}
Trapped singly-charged ions can crystallize as a result of laser cooling. The emerging structure depends on the number of particles and on the geometry of the trapping potential. In linear multipole radiofrequency traps, the geometry of the radial potential can lead to the formation of single-ring structures. We analyse the conditions and stability of single rings as a function of the number of poles. The rings form tubes for a large number of ions and sufficiently small trap aspect ratios. In these structures the arrangement of the ions corresponds to a triangular lattice folded onto a cylinder. The stability of the tubular structures is numerically studied for different lattice constants and their normal mode spectrum is determined. 
\end{abstract}

\maketitle

\section{Introduction}

Singly-charged particles in an external trapping potential can form ordered structures at sufficiently low temperatures, for which the interaction energy exceeds the thermal energy \cite{Dubin1999,Wigner}. The structures emerge from the balance between the Coulomb repulsion and the confining force, and their form can be thus controlled by changing the geometry of the trap \cite{Dubin1999}. Long-range order in structures of thousands to millions of ions were experimentally demonstrated in Penning and linear Paul traps \cite{Bollinger_Science,Drewsen}.

While most works have focussed on harmonic traps, or better said, potentials which can be approximated by harmonic ones, most recently research has focussed on dynamics and crystallization in anharmonic potentials formed by multipole traps \cite{Gerlich2007}.  
Recent works have studied the formation of ordered structures \cite{Okada2007, Okada2009, Calvo_PRA_2009, Champenois2010, Yurtsever_PRA_2011, Marciante2012} and crystallization in octupole traps has been observed by means of laser cooling \cite{Okada2009}. These studies show that ions in multipole traps tend to form hollow structures at very low temperatures. In linear multipole traps the structures take the form of either rings, tubes, or cylinders formed by several layers of tubes. The degree of ordering in these configurations depends on the trap parameters and on the number of ions: Numerical simulations indicate that both chirality and localized defects can appear as a result of a mismatch between the number of ions and the ``magic number'' that would correspond to a perfect pattern for the given trap parameters \cite{Yurtsever_PRA_2011}. Applications are being discussed for metrology and quantum simulators \cite{Champenois2010}. This perspective further motivates a detailed characterization of these systems.

In this work we theoretically analyze the conditions under which single rings form in a multipole linear trap, like the one sketched in Fig. \ref{fig:trap}, and study their stability as a function of the number of poles. The conditions for stability of a single, two, or multiple rings are determined.  The study is extended to a large number of ions for the purpose of identifying the parameter regime for which ion tubes, namely, an ordered array of single rings, are stable. The eigenmodes and eigenfrequencies of the tubes are then evaluated. 

The article is organized as follows: In Section \ref{sec:multipotential} we introduce the effective potential for linear multipole traps. Section \ref{sec:equilibrium} is devoted to the analysis of the stability of single rings in such potentials. In Section \ref{sec:tubes} the conditions for the stability of tubular structures are investigated and the normal modes are determined. Finally, in Section \ref{sec:conclusions} we summarize the main results and discuss some outlooks to this work. The appendix provides further details relevant to the discussion in Sec. \ref{sec:equilibrium}.

\section{The model}
\label{sec:multipotential}

In this section we introduce the basic Hamiltonian that determines the dynamics of $N$ singly-charged particles confined by a linear multipole trap. Some preliminary considerations on the charge distribution in presence of an anharmonic potential are made. 

\subsection{Hamiltonian}

The particles are all assumed to possess the same charge $q$ and mass $m$. They mutually repel via the Coulomb interaction and are subject to a trapping potential which we denote by $V_{\rm trap}({\bm r})$ and is a function of the position ${\bm r}$. At sufficiently low density the particles are distinguishable and we label them by $j=1,\ldots, N$, so that ${\bm r}_j$ and ${\bm p}_j$ are the canonically-conjugated position and momentum of ion $j$. The energy of the charges is given by Hamiltonian 
\begin{equation} \label{eq:hamilton-function}
 H = \sum_{j=1}^N \frac{\bm p_j^2}{2m} + \sum_{j=1}^N V_{\rm trap}(\bm r_j) + \frac{q^2}{4 \pi \epsilon_0} \sum_{j=1}^N \sum_{l=j+1}^N \frac{1}{\left| \bm r_j - \bm r_l\right|}\,,
\end{equation}
with $\epsilon_0$ the vacuum permittivity. 

\begin{figure}[hbt]
\includegraphics[width=0.6\columnwidth]{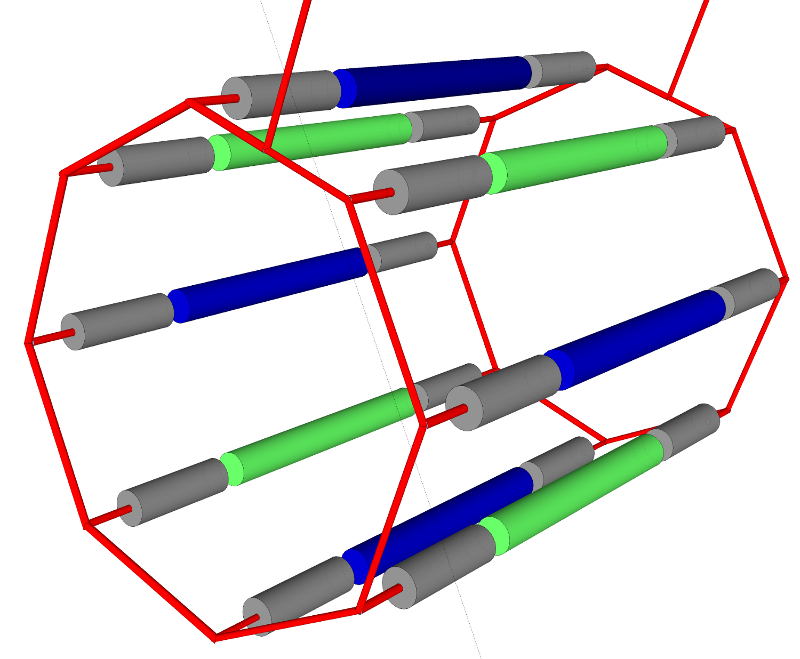}
 \caption{\label{fig:trap} Schematic representation of a linear octupole radiofrequency trap. The trap consists of eight parallel rods that are situated on a ring with equal radial spacing. Each rod is separated into three segments. An oscillating voltage is applied to the adjacent parts of the central segments of the electrodes and a static voltage to the outer segments of the electrodes \cite{Okada2007}.}
\end{figure}

The potential $V_{\rm trap}({\bf r})$ is generated by a radio-frequency trap. The trap consists of a set of $2k$ parallel electrodes that are equally spaced on a cylinder with radius~$r_0$, as shown in Fig. \ref{fig:trap} for $2k=8$. To provide the confinement along the $z$-axis, the electrodes can be split into three segments, and a positive static voltage is applied to the endcaps of the electrodes \cite{Okada2007}. For small voltages and in the central region of the trap, the resulting static potential can be approximated by the equation
\begin{equation}
 V_{\rm st}({\bm r}) = \frac 1 2 m \omega_z^2 \left(z^2 - \frac 1 2 r^2 \right) ,
\end{equation}
with $\omega_z$ the axial trapping frequency, which depends on the static voltage applied, while $r = \sqrt{x^2 + y^2}$ is the radial coordinate and $z$ the position along the trap axis. The radial confinement is generated by a rapidly oscillating voltage, which is applied to the central part of each electrode. Such setup creates a potential that in cylindrical coordinates takes the form~\cite{Ghosh1995,Gerlich2007}
\begin{equation}\label{rfpotential}
 \Phi_{\rm rf}(r,\theta) = V_0 \left( \frac{r}{r_0} \right)^k \cos(k \theta) \cos\left(\Omega t\right),
\end{equation}
where $\Omega$ is the angular frequency of the oscillation, $\theta$ is the azimuthal angle, $t$ is the time, and $V_0$ is a real positive scalar with the dimensions of an energy, such that the amplitude difference between two adjacent electrodes is~$2 V_0$. For $k>2$ a general solution of the equations of motion is not known. Nevertheless, in some cases the orbits of a single ion in potential \eqref{rfpotential} can be well approximated by the superposition of a slow motion along a smooth curve and a fast oscillation about it at much smaller amplitude, as it can be done in the stability region of a Paul trap. This is verified provided that the relative variation of the local electric field, seen over the amplitude of the oscillations of the driven motion, is sufficiently small, namely, when $\eta \lesssim 0.3$, where $\eta$ is the so-called adiabaticity parameter defined as
\beq
\eta = \frac{2q}{m \Omega^2}\frac{|({\bf E}\cdot \nabla){\bf E}|}{|{\bf E}|}\,,
% \left( \left|\frac{\partial E_x}{\partial x}\right| + \left|\frac{\partial E_y}{\partial y}\right| + \left|\frac{\partial E_z}{\partial z}\right| \right)
\eeq
and ${\bf E}$ is the local field generated by the oscillating potential \cite{Teloy1974, Champenois2009}. In this limit the secular motion is subject to an effective conservative force associated to the electrostatic potential usually denoted by pseudopotential, which in this case reads \cite{Champenois2009}
\begin{equation}
\label{eq:eff potential}
 V_{\rm rf} ({\bm r})=
 % \frac{1}{4m}\left(\frac{q \, V_0 \,  k}{\Omega \, r_0^k}\right)^2 \, r^{2k-2} \,.
 \frac{1}{4} V_{\rm rf}^{(0)}  \, r^{2k-2} \,,
 \end{equation}
with $V_{\rm rf}^{(0)}=q \, V_0 \,  k/(m\Omega \, r_0^k)$. The effective trap potential, which confines a particle in all three directions of space, is then given by
\beq \label{eq:trap-pot}
V_{\rm trap}({\bm r}) = V_{\rm rf} ({\bm r})+ V_{\rm st}({\bm r}) \,,
\eeq
and is the potential we consider in Eq. \eqref{eq:hamilton-function}. It neglects any influence of mirror charges on the conducting boundaries of the electrodes, which is a good description if the ions are sufficiently far away from the electrodes~\cite{Dubin1999}.

Potential $V_{\rm trap}({\bm r})$ can be written as a sum of two contributions, one depending on the radial coordinate $r$ and the other on the axial coordinate $z$. While in a quadrupole trap ($k$=2) the pseudopotential is harmonic and its minimum is at the trap center (which here coincides with the origin of the coordinates), in a multiple trap, for $k>2$, the center of the trap is unstable and the potential is minimum along a cylinder centered at the trap axis with radius \cite{Okada2007,Champenois2009}
\beq
\label{r:min}
r_{\rm min}=\left[\frac{m \, \omega_z^2}{(k-1)V_{\rm rf}^{(0)}}\right]^{1/(2k-4)}\,.
\eeq
When more than one ion is inside the trap, the Coulomb interaction must be taken into account to determine the equilibrium density. 

\subsection{Charge distribution at equilibrium for weak correlations}

The distribution of charges at equilibrium in a multipole trap has been numerically evaluated in a series of works, which have reported the appearance of hollow structures \cite{Okada2007, Okada2009, Calvo_PRA_2009, Champenois2010, Yurtsever_PRA_2011, Marciante2012} . This phenomenon can already be  understood by calculating the charge density distribution of a weakly correlated plasma \cite{Champenois2009}. For particle numbers of the order of hundreds and more, the thermal equilibrium states can be described by a Gibbs distribution. The particle density, which we denote by $n( \bm r)$, can be approximated by the single-particle Boltzmann distribution in the limit in which the Debye length $\lambda_D$ is small with respect to the plasma dimensions~\cite{Dubin1999}. The density takes the form
 \begin{equation} \label{eq:boltzmann}
  n(\bm r) \simeq A \exp\{- \beta [V_{\rm trap}(\bm r) + q \, \Phi_p(\bm r)]\}.
 \end{equation}
where $\beta=1/k_BT$, with $k_B$ Boltzmann constant and $T$ the temperature, $A$ is a scaling constant warranting that  $\int n(\bm r) d^3r = N$, and $\Phi_p$ is the electric potential created by the charge distribution itself, satisfying the Poisson equation
 \begin{align} \label{eq:poisson}
  \Delta \Phi_p(\bm r) = - \frac{q}{\epsilon_0} n(\bm r)\,.
 \end{align}
Potential $\Phi_p$ is assumed to vanish at the conducting surfaces of the trap, and infinitely far away from the charge distribution (these boundary conditions allow for neglecting the mirror charges only when the conducting surfaces are taken to be far away from the plasma). When the potential is harmonic the density is uniform, $n(\bm r)=n_0$, and the thermodynamic properties are solely characterized by the so-called plasma parameter $\Gamma=q^2n_0^{1/3}/(4\pi\epsilon_0k_BT)$. The Debye length takes the form $\lambda_D = \sqrt{k_B T \epsilon_0/(q^2 n_0)}$, and is larger than the interparticle distance when $\Gamma<1$, since $n_0\lambda_D^3=(4\pi\Gamma)^{-3/2}$.

In multipole traps the scenario is quite different. Solving  Eq. \eqref{eq:poisson} in presence of an anharmonic potential leads to non-uniform densities: the density first increases with the distance to the trap axis, reaches a maximum at a non-zero radius, and then rapidly drops to zero. We refer the reader to the appendix and Ref.~\cite{Champenois2009} for detailed calculations. This behaviour reflects the form of the radial dependence of the potential for multipole traps. One can define a plasma parameter $\Gamma(\bm r)=q^2n(\bm r)^{1/3}/(4\pi\epsilon_0k_BT)$, which results to be position dependent, and is hence smaller at the center of the trap, reaching a maximum at the radius where the density is maximum. This feature indicates that crystallization is first expected in the outer shells of the crystals, and only at sufficiently low temperatures the ions form ordered clusters in the entire trap volume where they are distributed~\cite{Calvo_PRA_2009,Yurtsever_PRA_2011}.
   
\subsection{Charge distribution at equilibrium for strong correlations}

For uniform distributions in three dimensions a singly-charged plasma exhibits  at $\Gamma\sim 174$ a first-order phase transition from a liquid to a crystal phase \cite{Dubin1999}. Similar to the case of harmonic traps, in anharmonic traps at sufficiently low densities and temperatures the ions crystallize. By this, we mean they arrange in structures in which each ion has a well-defined equilibrium position~$\bm r^{(0)}_j$ and performs small oscillations about it, whose amplitudes are much smaller than the interparticle distance. In an equilibrium configuration the net force over each ion vanishes, namely,
$$\left. \frac{\partial V}{\partial \bm r_j} \right|_{\bm r_j = \bm r_j^{(0)}} = 0\,,$$ 
where $V$ is the total potential, corresponding to the potential term in  Eq. \eqref{eq:hamilton-function}, which we here rewrite as
 \begin{equation} \label{eq:pot-energy}
 V= a_1 \sum_{j=1}^N r_j^{2k-2} + a_2 \sum_{j=1}^N \left(z_j^2 - \frac 1 2 r_j^2\right) + a_3 \sum_{j<l}\frac{1}{\left| \bm r_j - \bm r_l\right|},
\end{equation}
with 
\beq
a_1=\frac{1}{4} \,V_{\rm rf}^{(0)}  , \quad
a_2=\frac{1}{2} m \omega_z^2 \, , \quad
a_3 = \frac{q^2}{4 \pi \epsilon_0} \, .
\eeq
The set of positions $\{\bm r^{(0)}_j\}$ describes a stable configuration if the potential there evaluated is a local minimum, which corresponds to the condition that the Hessian of the potential energy at the positions $\bm r^{(0)}_j$ is positive-definite. We note that, in general, there will be several possible local minima for a given number of system parameters. Unless otherwise stated, we will discuss the configuration that corresponds to a global minimum of the potential energy.

\section{Ion rings}
\label{sec:equilibrium}

In this section we analyze the stability of single rings which, for a given number of ions,  are formed at the center of the trap for a sufficiently large aspect ratio between the radial and the axial trapping potential. 

In order to study the problem, we first introduce dimensionless coordinates, and denote by $\bm{\tilde r} = \bm r /\bar r$ the dimensionless position, with $\bar r=(a_1/a_3)^{-1/(2k-1)}$. The dimensionless potential $ \tilde V$ is obtained by rescaling potential $V$ in Eq. \eqref{eq:pot-energy} by the quantity  $a_3/\bar r$ and reads 
\begin{equation} \label{eq:pot-energy-ndim}
 \tilde V = \sum_{j=1}^N \tilde r_j^{2k-2} + c \sum_{i=1}^N \left(\tilde z_j^2 - \frac 1 2 \tilde r_j^2\right) + \sum_{j<l}\frac{1}{\left| \bm{\tilde r}_j - \bm{\tilde r}_l\right|},
\end{equation}
where the parameter
\beq
c = a_2 \bar r^3/a_3 
\eeq
depends on the relative strength of each contribution. Equation \eqref{eq:pot-energy-ndim} shows that every cluster is determined by three independent parameters, which can be chosen to be $N$, $k$, and $c$. In the rest of this section we discuss the equilibrium configurations and omit the tilde (but always refer to the dimensionless coordinates, unless otherwise stated). 

The equilibrium configurations are found by numerically minimizing the potential energy in Eq.~(\ref{eq:pot-energy-ndim}) using simulated annealing~\cite{Ingber1993,Coleman1996}. Before we start, we remark that single ring structures in linear multipole traps have been reported in Refs. \cite{Okada2007,Champenois2010,Marciante2012}. In Ref. \cite{Champenois2010} the conditions for stability of a single ring were derived and in Ref. \cite{Marciante2012} numerical simulations based on molecular dynamics were reported. The results shown below are in agreement with these studies and complement them since they provide a systematic characterization of the transition from a single to a double ring by means of a closed set of equations.

\subsection{Single rings}

\begin{figure}
 \subfloat[]{\includegraphics[width=0.27\textwidth]{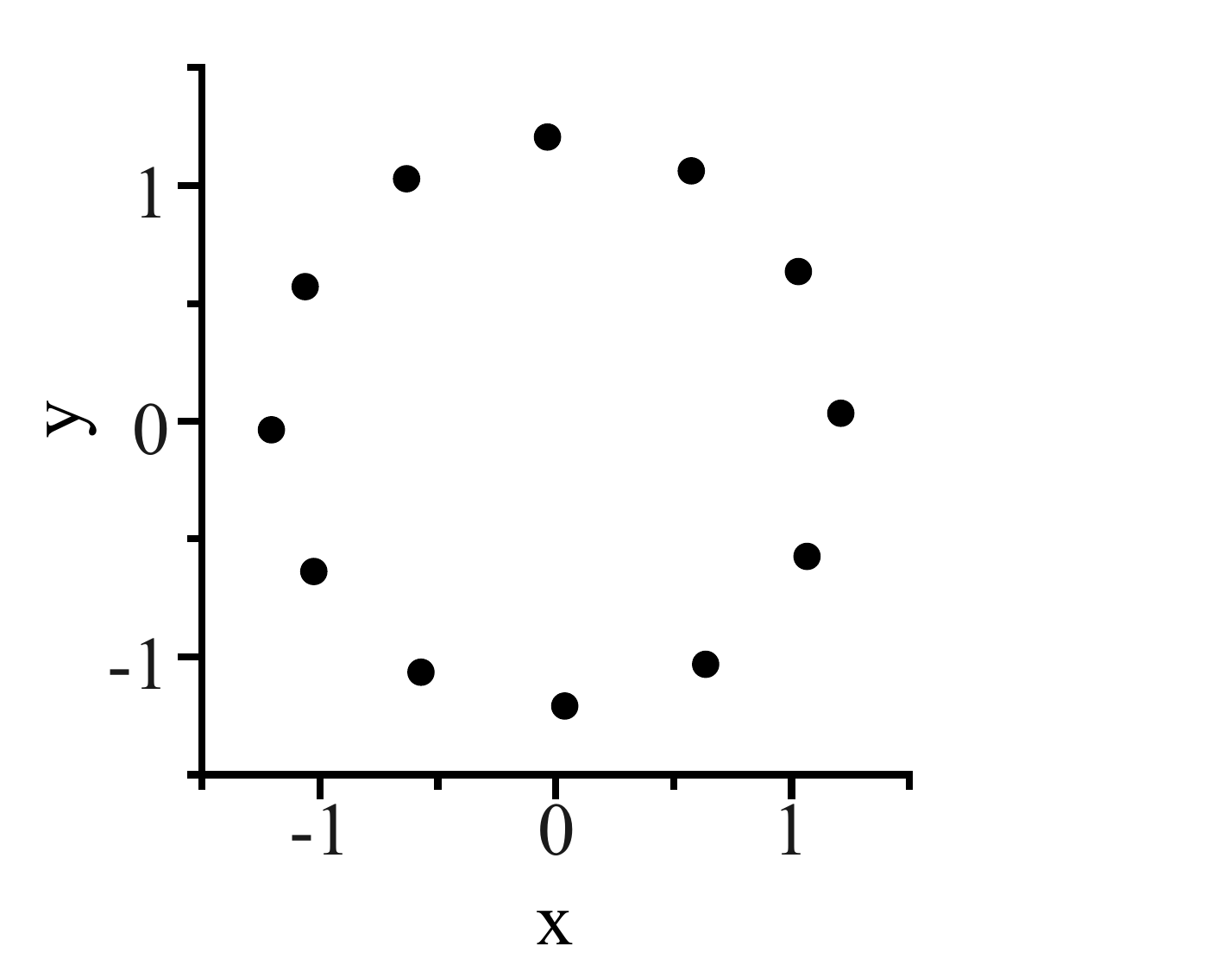}} 
 \subfloat[]{\includegraphics[width=0.27\textwidth]{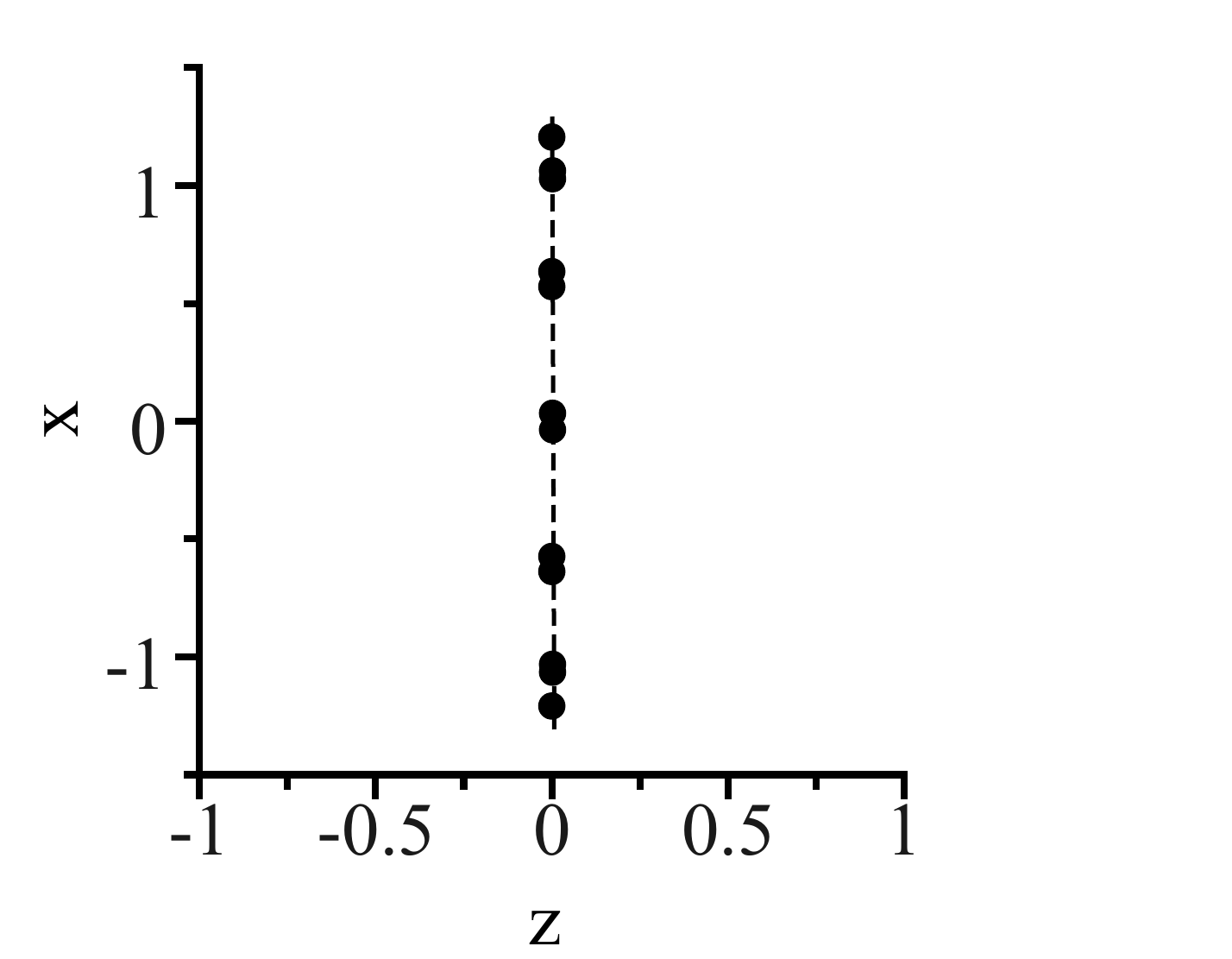}} \\
 \subfloat[]{\includegraphics[width=0.27\textwidth]{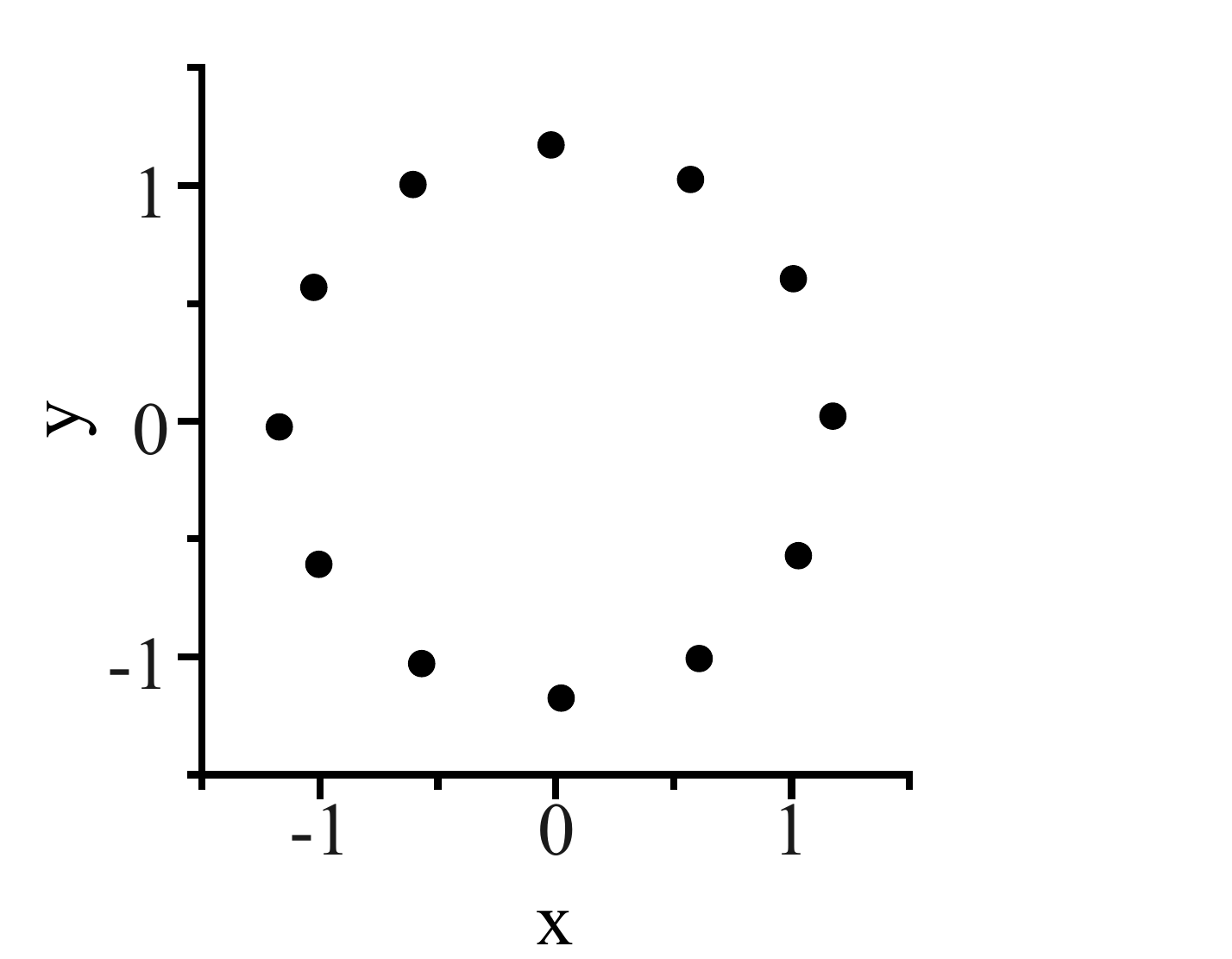}} 
 \subfloat[]{\includegraphics[width=0.27\textwidth]{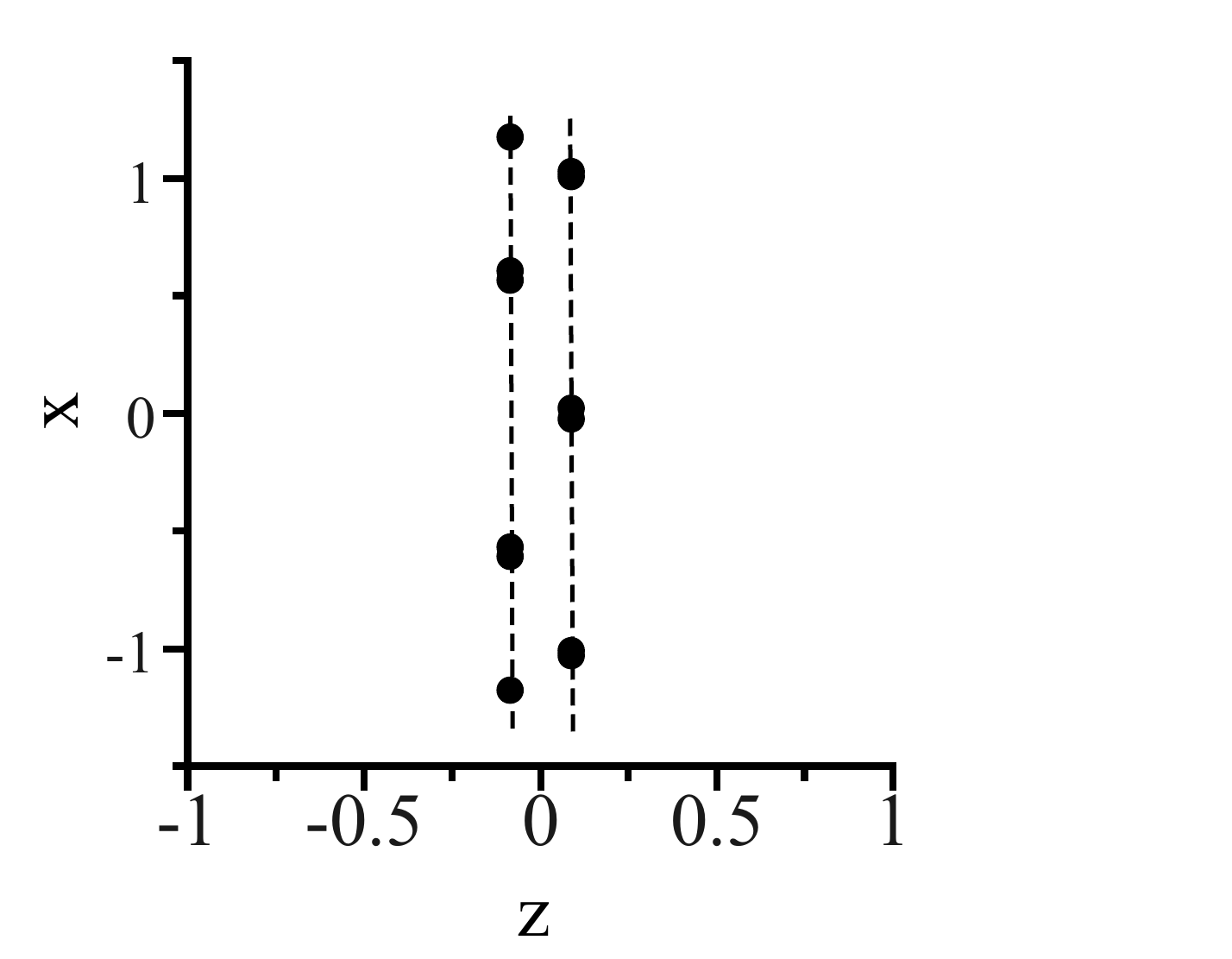}} \\ 
 \caption{Equilibrium configurations of $N=12$~ions in an octupole ($k=4$) trap, determined by simulated annealing. The panels on the left show the charge distribution in the radial, $x-y$, plane for different values of $c$, the panels on the right the corresponding distribution on the plane $x-z$. In (a)-(b) the value $c=10$ was taken. The plots in (c)-(d) are evaluated for $c=8.5$ and exhibit a double ring structure.} 
 \label{fig:linear-zigzag}
\end{figure}

We first focus on the existence of single rings in the plane $z=0$ of the multipole trap. Given a number of ions $N$ in a multipole trap with $k>2$, one can always find a sufficiently large value of $c$ for which the ions arrange on a single ring at the plane $z=0$, where the ring radius $R$ depends on the three parameters $N,k,c$. Some configurations are shown in Fig. \ref{fig:linear-zigzag}. Analytical expressions for the radius have been derived in \cite{Champenois2010} and will be reported later on for studying the transition to the double ring visible in Figs. \ref{fig:linear-zigzag}(c) and~(d).  

Few remarks are here in order. For sufficiently large aspect ratios between the radial and the axial potentials (corresponding to large values of $c$), in a linear Paul trap single-ring structures are a global minimum of the potential only for $1<N\leq5$ \cite{footnote}. In multipole linear traps, instead, these configurations are found for any number $N$ provided that $c$ is sufficiently large. They result from the specific shape of the anharmonic potential in Eq.~(\ref{eq:trap-pot}), having a minimum shifted away from the trap center. Indeed, even for a single ion the radius $r_{\rm min}$, Eq. \eqref{r:min}, increases as $c$ is increased either by decreasing the radial confinement (and thus $a_1$) or by rendering the axial confinement steeper (increasing $a_2$).

\subsection{From a single to a double ring}

We now analyze the ions' configuration as $c$ is decreased keeping $k$ and $N$ constant. In this case, a single ring in the plane $z=0$ is stable until $c$ reaches a critical value $c_{\rm crit}$, below which the ions' distribution is no longer planar. An analogous behaviour is observed if $c$ is kept fixed but the number of ions $N$ is increased: in this case the transition to a three-dimensional structure is observed when $N$ exceeds a value $N_{\rm crit}$ which depends on $k$ and $c$.  

\begin{figure}
 \includegraphics[width=0.8\columnwidth]{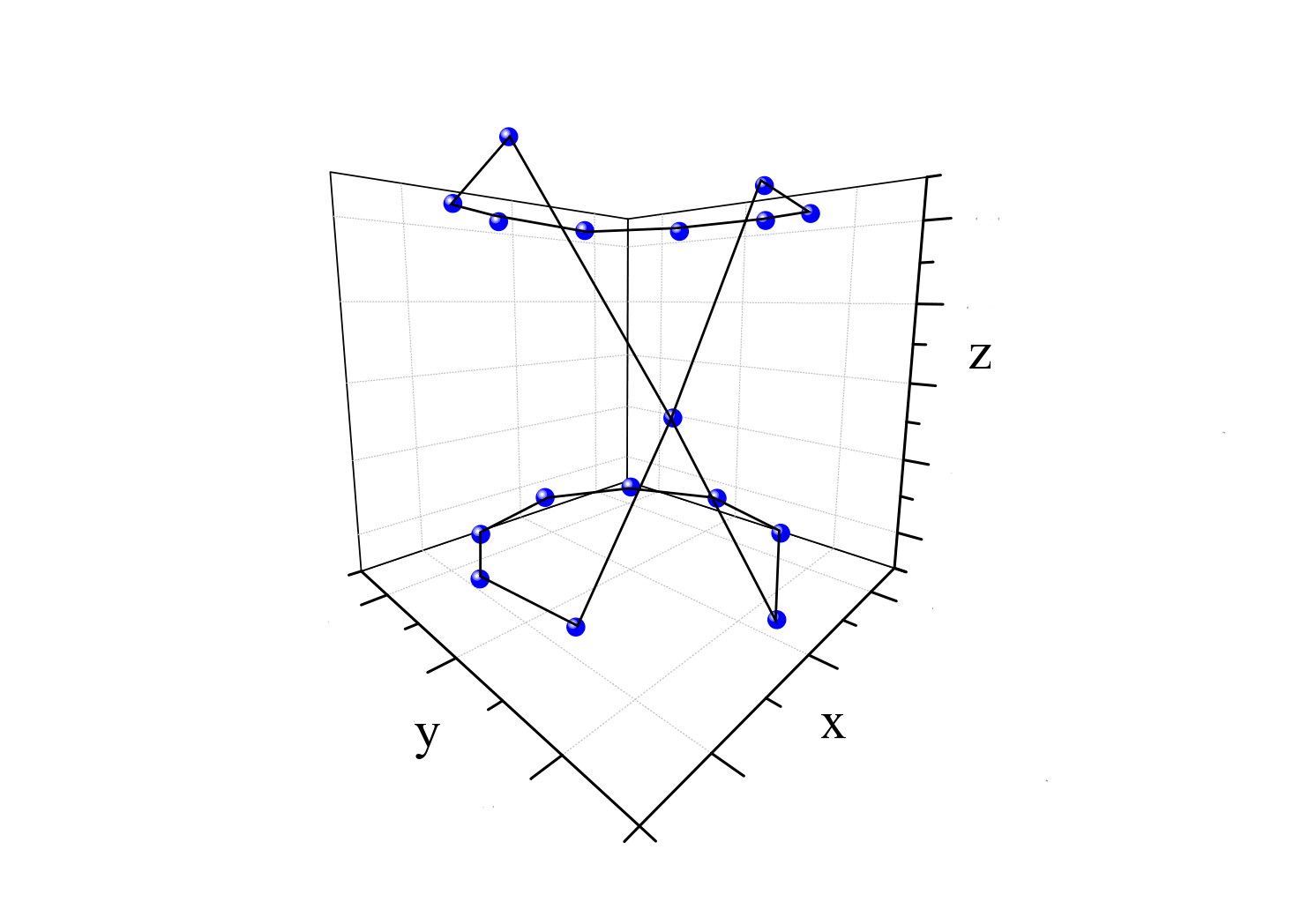}
 \caption{Equilibrium configuration of 17 ions in a linear multipole trap for $k=4$ and $c=10$, determined by simulated annealing. The ions form two rings with equal number $M=8$ of ions, while one ion is located in the plane $z=0$ between both rings, forming a localized defect. }
 \label{fig:odd}
\end{figure}

Be now $N$ constant and $c$ below (but sufficiently close to) $c_{\rm crit}$: For an even total number of ions, $N=2M$, the ions arrange in two rings with the same number of ions and the same radius, as shown in Fig. \ref{fig:linear-zigzag}(c)-(d). In Fig. \ref{fig:linear-zigzag}(d) one observes that the structure seen from the side takes a zigzag shape. This kind of shape was also reported in Refs. \cite{Champenois2010,Marciante2012}. For an odd number of ions, $N=2M+1$, the ions form two rings with equal number $M$ of ions, while one ion sits in the plane $z=0$ between both rings, forming a localized defect, as shown in Fig.~\ref{fig:odd}. 

\begin{figure}
 \subfloat{\includegraphics[width=0.27\textwidth]{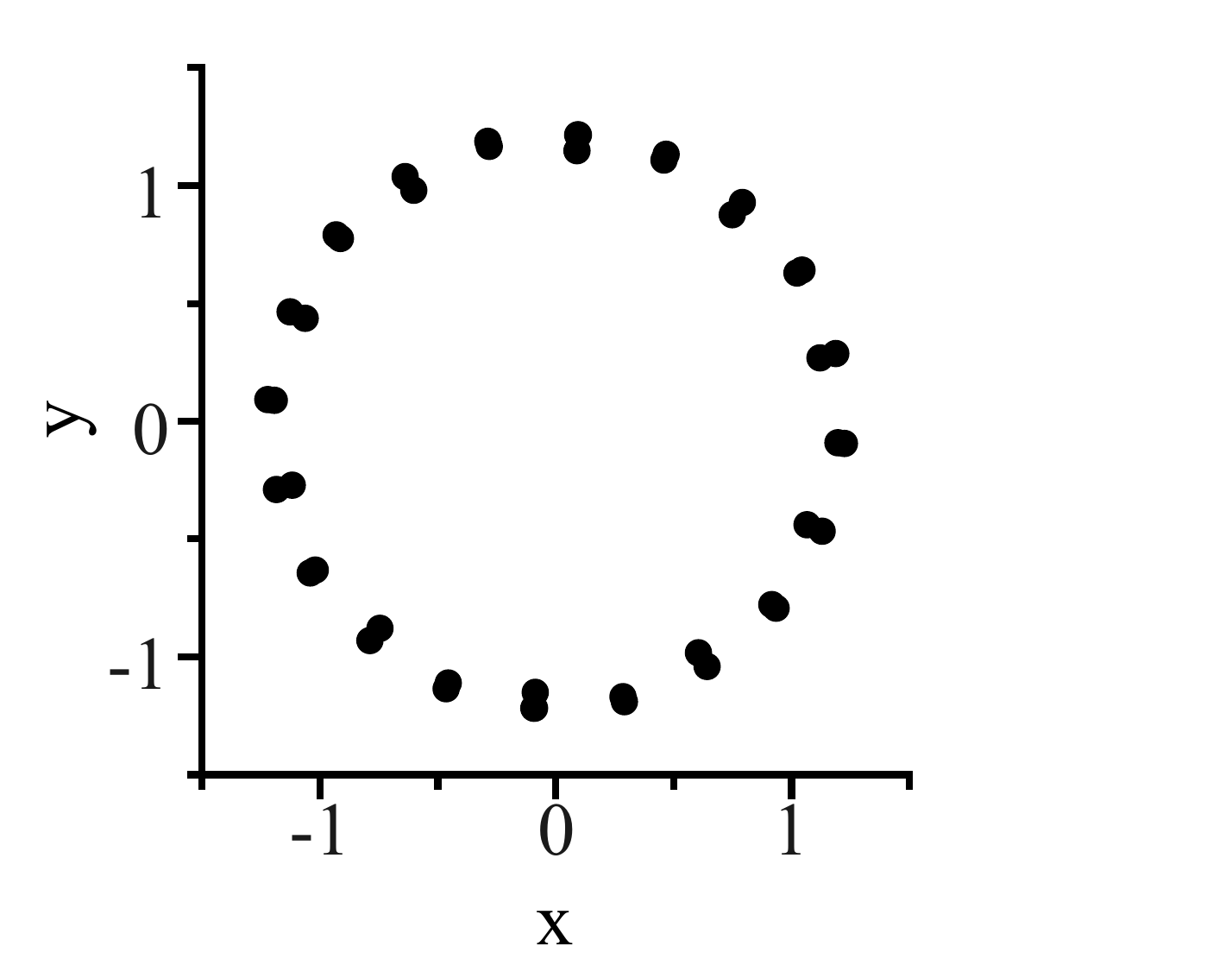}} 
 \subfloat{\includegraphics[width=0.27\textwidth]{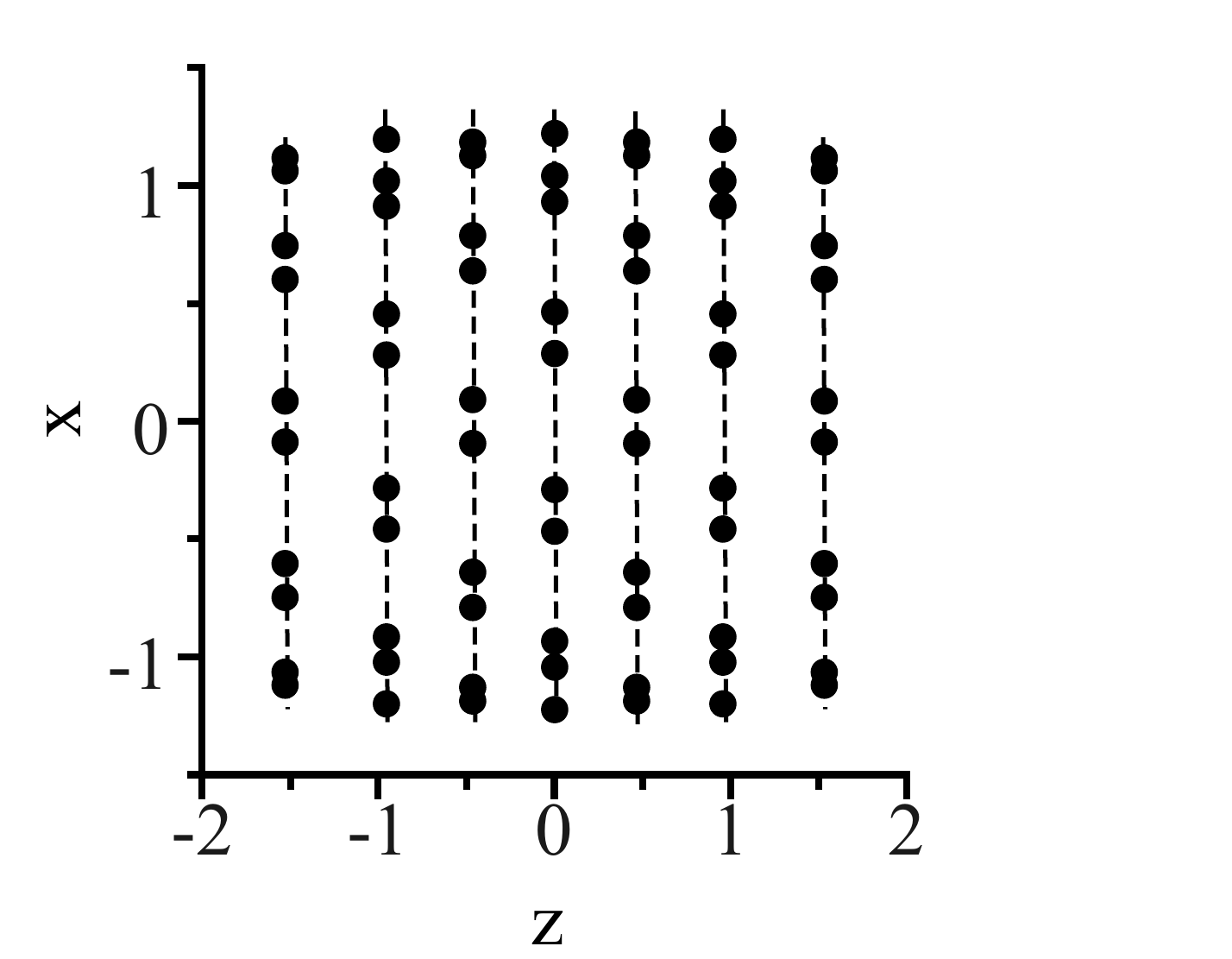}} 
 \caption{Equilibrium configuration of 70~ions in an octupole trap~($k=4$) for $c=4$, corresponding to a tubular arrangement. The positions have been found by simulated annealing. The inner rings are composed by 12 ions per ring, while at the two edges the rings are composed by 10 ions each.} \label{fig:clusters}
\end{figure}

By further decreasing $c$, parameter regimes are found for which the ions arrange in multiple rings forming a cylinder, as displayed in Fig.~\ref{fig:clusters}. In particular, all rings have the same number of ions and approximately the same radius, except for the rings at the edges that have a smaller radius and also a smaller number of ions. 

In the following we derive an analytical expression for the instability of a single ring composed by an even number of ions. Close to the instability point, the following ansatz describes the ions' equilibrium positions either in the single or double ring configuration:
\beq \label{eq:double-ring}
\left\{
\begin{array}{l}
x_j = R \cos ( 2\pi j/ N ) \\
y_j = R \sin ( 2\pi j/ N ) \\
z_j = \left( -1 \right)^{j} d/2
\end{array}
\right.\,
\quad
  j = 1,\ldots ,N\,,
\eeq
where $R$ is the (dimensionless) ring radius and $d/2$ is the (dimensionless) distance of the ions from the plane $z=0$. Therefore, if $d=0$, all ions are located in the plane $z=0$ forming a single ring of radius $R$, while for $d\neq 0$, the ions form two rings, each consisting of $N/2$ ions, with radius~$R$ and at distance~$d$ from each other. The potential energy of a double-ring configuration depends on the radius~$R$ and distance~$d$ between rings, and is obtained by using Eqs.~(\ref{eq:double-ring}) in Eq.~(\ref{eq:pot-energy-ndim}). 

We now derive the relations that allow us to find the critical value $c_{\rm crit}$ for which the transition from a single ring to a double ring occurs. If a single ring with radius~$R=R_0$ is stable, then the potential energy will have a minimum at~$d=0$. This implies that the first derivatives of the potential energy with respect to $R$ and $d$ have to vanish,
\begin{align} \label{eq:dring-zero-force}
 \frac{\partial V}{\partial R} \Bigg|_{(R_0,0)}= 0 \qquad
 \frac{\partial V}{\partial d} \Bigg|_{(R_0,0)} = 0\,,
\end{align}
where the second condition always holds for symmetry reasons. The second derivatives must satisfy:
\begin{align}
 \frac{\partial ^2 V}{\partial R^2} \Bigg|_{(R_0,0)} > 0 \qquad \frac{\partial ^2 V}{\partial d^2} \Bigg|_{(R_0,0)} > 0\,,
\end{align}
while the mixed derivative $\partial^2 V/(\partial R \partial d)$ vanishes at \mbox{$d=0$}. 

\begin{figure}
 \subfloat{\includegraphics[width=0.8\columnwidth]{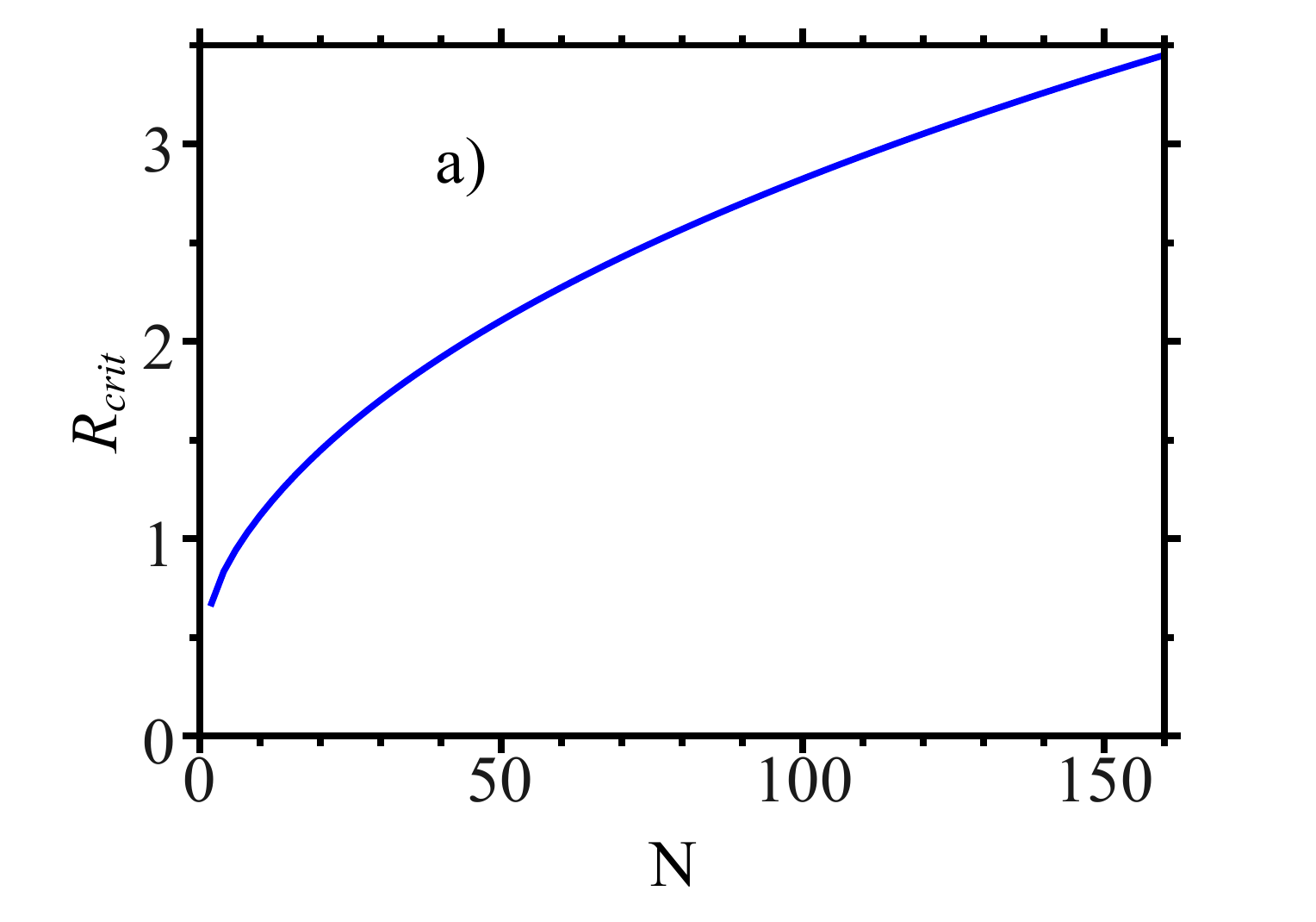}}\\
 \subfloat{\includegraphics[width=0.8\columnwidth]{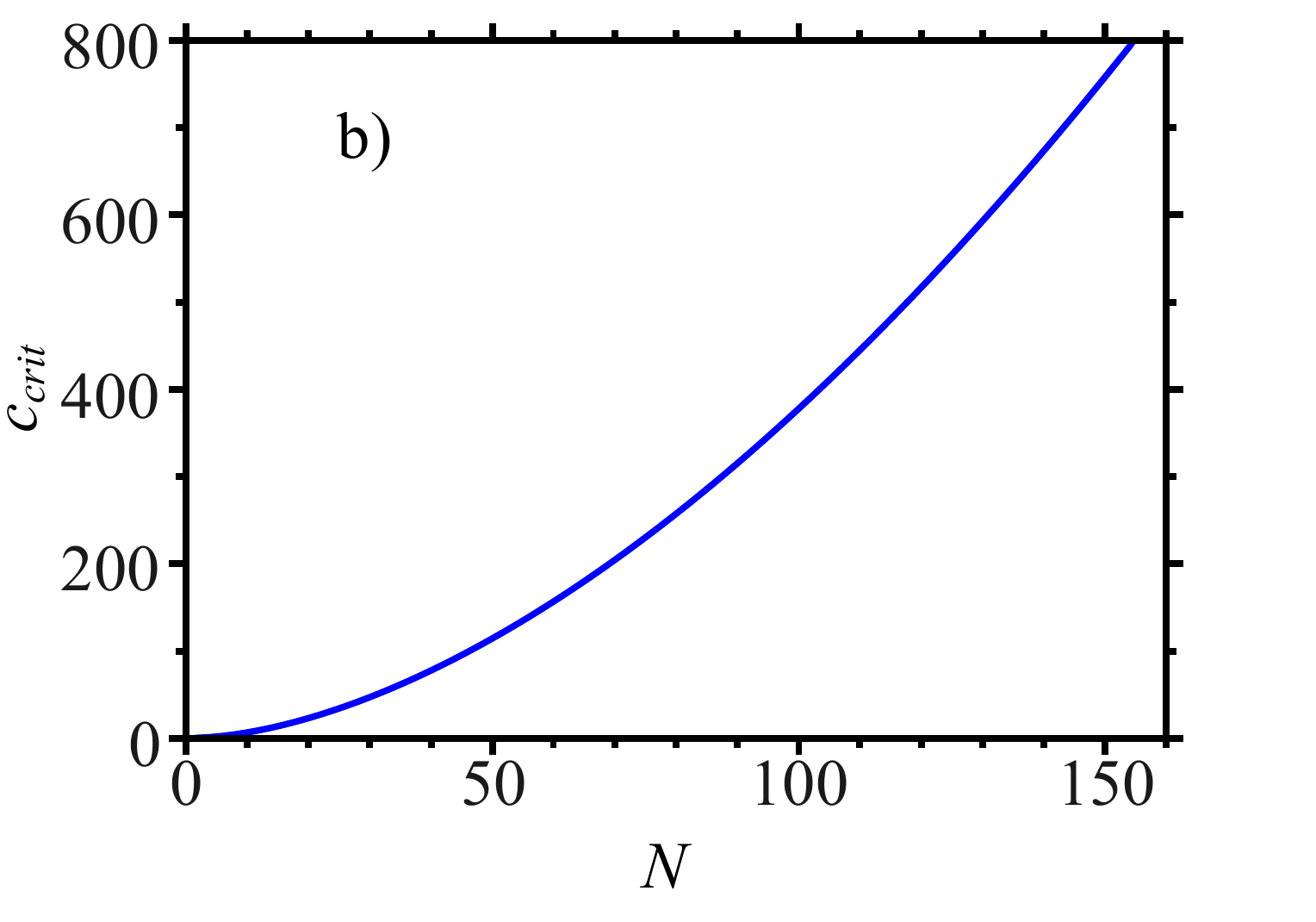}}
 \caption{\label{fig:c-r-crit} Critical parameters $R_{\rm crit}$ (a) and $c_{\rm crit}$ (b) at the transition point from a single to a double-ring structure as a function of the particle number for an octupole trap~($k=4$). The curves are evaluated from the analytical expressions in Eqs. \eqref{c:crit}-\eqref{R:crit}.}
\end{figure}

At the point where a single ring becomes unstable and splits into a double ring, the second derivative with respect to $d$ changes its sign crossing a zero,
\begin{align} \label{eq:dring-2nd-der}
 \frac{\partial ^2 V}{\partial d^2} \Bigg|_{(R_0,0)} = 0\,.
\end{align}
To find the point where a single ring becomes unstable, we solve Eqs. (\ref{eq:dring-zero-force}) and (\ref{eq:dring-2nd-der}) and find that the parameter~$c_{\rm crit}$ reads
\begin{align}
\label{c:crit}
 c_{\rm crit} &= \frac{1}{8} S_3(N)  \left[\frac{2S_1(N) + S_3(N)}{16(k-1)} \right]^{-3/(2k-1)}\,, \end{align}
 where
\begin{align}
 S_1(N) &= \sum_{j=1}^{N-1} \frac{1}{\sin \left( \frac{\pi j}{N} \right)} \,, \\
 S_3(N) &= \sum_{j=1}^{N/2} \frac{1}{\sin^3 \left( \frac{\pi (2j-1)}{N} \right)} \, .
\end{align}
The corresponding value of the ring radius reads
\beq
 R_{\rm crit} = \left[\frac{2 S_1(N) + S_3(N)}{16(k-1)} \right]^{1/(2k-1)}\,.
\label{R:crit}
\eeq
Figure~\ref{fig:c-r-crit} displays the parameters $c_{\rm crit}$ and $R_{\rm crit}$ as a function of the particle number $N$ in an octupole trap~($k=4$). For large particle numbers, the parameter~$c_{\rm crit}$ scales as $c_{\rm crit} \sim N^{6(k+1)/(2k-1)}$, while $R_{\rm crit} \sim N^{3/(2k-1)}$.  These behaviours show that in a quadrupole trap the critical radius scales linearly with the particle number, while for high orders of the multipole potential, $k \gg 1$, the critical radius reaches a constant value. In fact, for large values of~$k$, the multipole potential approaches a box potential of the form 
\begin{align}
 \lim_{k \to \infty} V_{\rm trap} = 
 \begin{cases}
 c  \left( z^2 - \frac{1}{2} r^2 \right) ,& 0 \leq r < 1 \\
 \infty, & r>1,
\end{cases} .
\end{align}
which has thus maximal radius $R_0=1$ (in dimensionless coordinates). 
Note that Eq. \eqref{R:crit} exhibits a different dependence on $N$ than the critical radius derived in Ref. \cite{Champenois2010}. This is due to the different conditions which were considered.  In \cite{Champenois2010} the axial trap frequency is fixed.  Here, we assume that the radial confinement is kept constant, while $c$ is varied by changing the axial trap frequency: Correspondingly, Eq. \eqref{R:crit}  (once rescaled by $\bar{r}=(a_3/a_1)^{1/(2k-1)}$) is reported in terms of $a_1$, $a_3$, $k$, and $N$. 

\subsection{Discussion}

The transition from a single to a double ring shares several analogies with the linear-zigzag transition observed  in a linear Paul trap \cite{Walther1992,Raizen, James2000}. Indeed, as for the linear-zigzag transition a symmetry is broken, which is the symmetry by reflection about the $z=0$ plane. However, in the case of a multipole  trap the radius of the distribution changes as the aspect ratio varies across the structural transition, and thus the density along the direction defined by the polar angle varies. An expansion close to the critical point shows that the distance $d$ increases like $\sqrt{\delta c}$ for $\delta c = c_{crit} -c>0$, while $R$ decreases linearly in $\delta c$. In addition, the effect of the curvature of the ring is relevant as long as the radius remains finite. Nevertheless, this system can be considered a platform where to test predictions related to quantum effects at the structural transition \cite{Shimshoni2011,Reznik2010}. For this purpose, an adequate analysis following the lines of the study in \cite{Morigi2004,Fishman2008} is required.    

Figure \ref{fig:clusters} shows an example of configuration which is found by further lowering the value of $c$. For a very large number of ions and weak confinement along the $z$ axis the ions form tubes, namely, uniformly spaced rings with the same radius and number of ions per ring. Their characterization is the subject of the next section.

\section{Ion tubes} \label{sec:tubes}

In this section we study the stability and dynamics of structures composed of multiple rings forming tubular structures. These structures have been first predicted in Ref.~\cite{Okada2007}. The arrangement of the ions in the central region of such tubes corresponds to a triangular lattice folded onto a cylinder. We note that previous studies based on molecular dynamics simulations have observed the formation of inner rings for systems with numbers of ions of the order of one hundred \cite{Okada2007} to one thousand \cite{Marciante2012}. Besides, the impossibility to fit any number of ions into a regular pattern for given trap parameters has been shown to lead to chiral structures or localized defects \cite{Yurtsever_PRA_2011}. In the following we will restrict to the study of tubular structures with no inner rings. Furthermore, we focus on the configurations with neither chirality nor defects. This means that we consider large ion numbers with very weak axial trapping for which single tubes are formed \cite{Yurtsever_PRA_2011}.

\subsection{Equilibrium structure}

\begin{figure}[ht]
 \includegraphics[width=0.65\columnwidth]{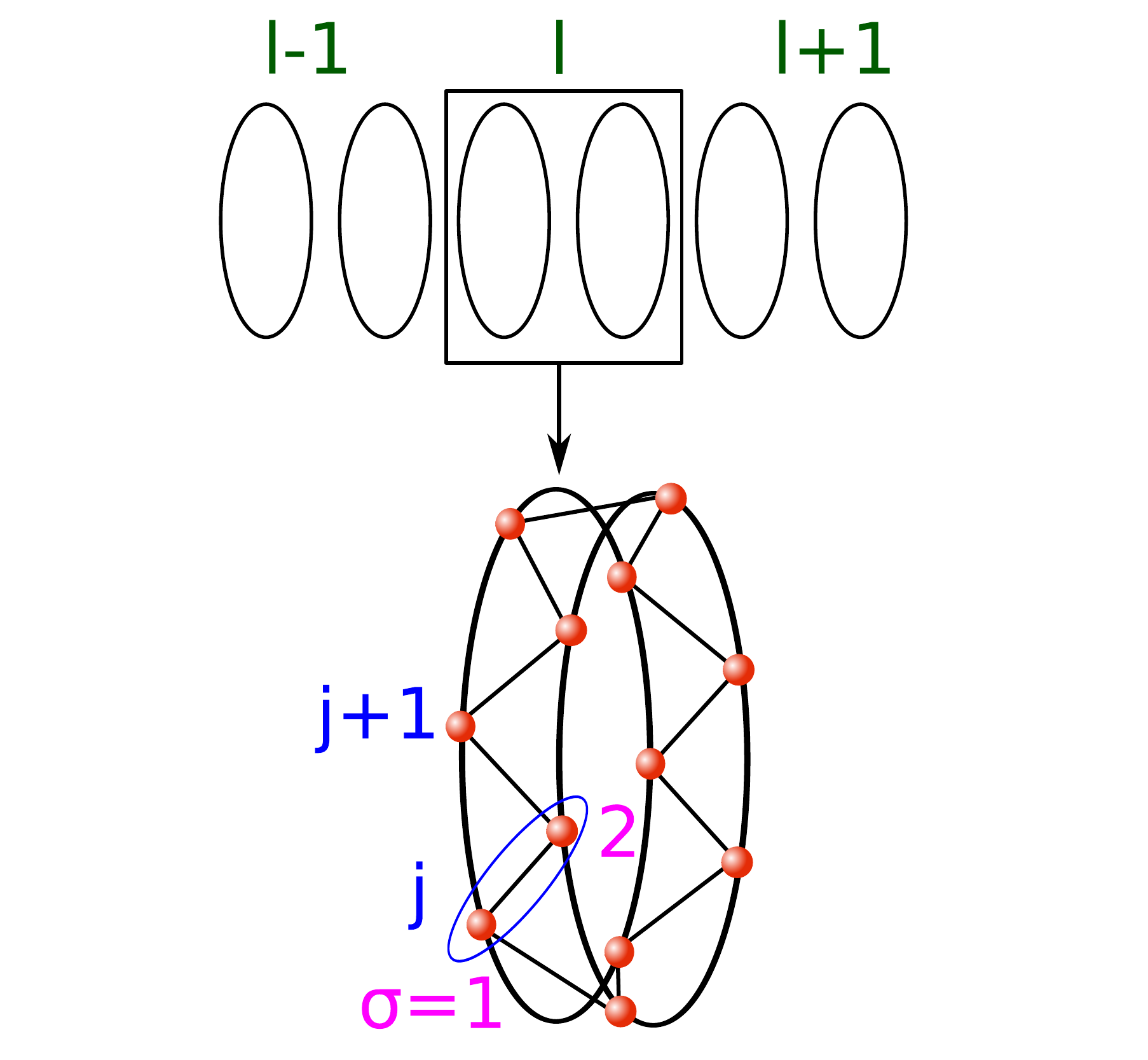}
 \caption{\label{Fig:WS} Top panel: sketch of the ring array at uniform distance $d/2$ along the $z$ axis. The lower panel shows the ions' distribution in two adjacent rings: this structure is repeated along the chain. An example of the chosen elementary cell is given by the pair of ions labeled with $\sigma$. The other labels are $j=0,\ldots,N_R-1$ for the number of cells in two adjacent rings and $l=0,\ldots,N_A-1$ for the pair of rings.}
\end{figure}

Let the tube be composed by $2 N_A$ rings, each with radius $R$ and $N_R$ ions per ring, and with distance  $d/2$ between adjacent rings. The tube is sketched in the upper panel of Fig.~\ref{Fig:WS}. Periodic boundary conditions in $z$ direction are assumed, taking that the trap confines the ions only in the radial direction, 
\begin{align}
V_{\rm trap}'=a_1 \sum_\alpha r_\alpha^{2k-2} \,,
\end{align}
while the axial confinement vanishes, $a_2=0$, so that the total potential $V'$ the ions feel reads 
$$V'= V_{\rm trap}' + \frac{q^2}{4 \pi \epsilon_0} \sum_{\alpha,\alpha'\neq\alpha} \frac{1}{\left| \bm r_\alpha - \bm r_{\alpha'}\right|}\,.$$ 
This assumption can be taken for a long tube in a linear multipole trap for which the boundary effects can be neglected. Under this assumption, the resulting structure possesses a discrete rotational symmetry by the angle $\phi_0 = 2 \pi/N_R$ about the $z$ axis, as visible in the lower panel of  Fig. \ref{Fig:WS}, and a discrete translational symmetry under the axial displacement $d$. These symmetries lead to the choice of  a unit cell composed by two ions, as shown in the lower panel of Fig.~\ref{Fig:WS}.  The total number of unit cells is $N_{\rm cells}=N_A \, N_R$, thus half of the total number of ions composing the tube.

In what follows we shall label the ions by the set of indices $\alpha = (j,l,\sigma)$: the index $j$ labels the cells along the azimuthal angle with $j=0,\ldots,N_R-1$, the index $l$ labels the cells along the $z$ direction with $l=0,\ldots,N_A-1$, while $\sigma=0,1$ labels the two ions in each unit cell, see also Fig.~\ref{Fig:WS}. The equilibrium positions of the ions are given by the vectors $ {\bm R}_\alpha \equiv {\bm R} (j,l,\sigma) $, which in cartesian coordinates read
\begin{align} 
  {\bm R}_\alpha =\begin{pmatrix}
  R \cos\left[ \phi_0 \left( j+ \sigma/2 \right) \right] \\
  R \sin\left[ \phi_0 \left( j+ \sigma/2 \right) \right] \\
  d \left( l + \sigma/2 \right)
 \end{pmatrix}.
\end{align}
Here, $R$ is found from the condition that the net force on each ion vanishes, and fulfills the equation
\begin{multline} \label{eq:trap-cond}
 \frac{a_1 R^{2k-1}}{q^2/(4 \pi \epsilon_0)}=\frac{2}{(2k-2)} \\
 \sum_{(j l \sigma) \neq (0 0 0)}\frac{\sin^2\left[ \frac 1 2 \phi_0 \left(j + \frac{\sigma}{2} \right)\right]}{\left\{4 \sin^2\left[ \frac 1 2 \phi_0 \left(j + \frac{\sigma}{2} \right)\right] + \left(\frac{d}{R}\right)^2\left( l + \frac{\sigma}{2} \right)^2 \right\}^{3/2}}.
\end{multline}
Note that the sum on the right-hand side converges for an infinitely long tube. 

If this configuration is stable,  at sufficiently low temperatures each ion performs small oscillations with a displacement~${\bm u}_\alpha$ from its equilibrium position~$\bm R_\alpha$, so that
\beq
 {\bm u}_{\alpha}={\bm r_\alpha} - {\bm R_\alpha} \,,
\eeq
where  ${\bm r}_\alpha$ indicates the position of the particle labelled by $\alpha$ and where periodic boundary conditions impose that ${\bm u}_{j,l,\sigma} = {\bm u}_{j+N_R,l+N_A,\sigma}$ (with ${\bm u}_\alpha\equiv {\bm u}_{j,l,\sigma}$). The harmonic approximation can be performed when the displacement is at all times much smaller than the interparticle distance. In this limit the potential energy can be approximated by its quadratic expansion  
\begin{align} \label{eq:harmomic-exp}
 V' \approx E_0 + \frac 1 2 \sum_{\alpha, \alpha'} \bm u^T_\alpha D(\alpha,\alpha') \bm u_{\alpha'}\,,
\end{align}
where $E_0 = V'(\{\bm R_\alpha\})$ is the energy of the equilibrium configuration, while $D$ is a matrix whose coefficients read
%\begin{widetext}
\begin{align}
 D_{\mu \nu}(\alpha,\alpha') = 
 \frac{\partial ^2 V'}{\partial {\bm r}_{\alpha,\mu} \partial {\bm r}_{\alpha',\nu} }\Bigl|_{\bm R_\alpha,\bm R_{\alpha'}} 
\qquad \mu,\nu \in \{x,y,z\}
\end{align}
%\end{widetext}
and all derivatives are evaluated at the equilibrium positions $\{\bm R_\alpha\}$. 

\begin{figure}[htb]
 \includegraphics[width=\columnwidth]{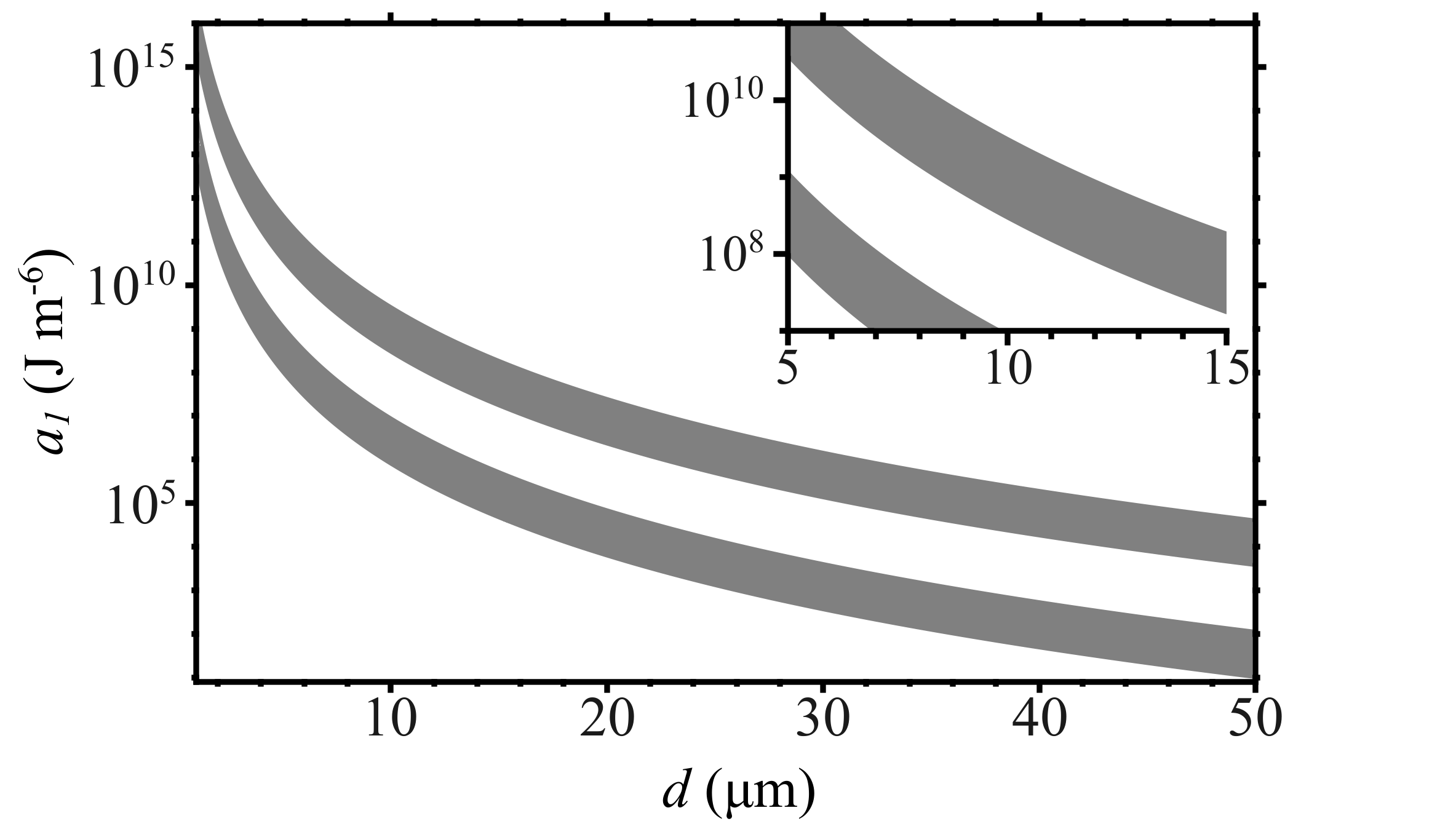}
 \caption{Stability diagram of ion tubes with six ions per ring ($N_R=6$) in an octupole trap ($k=4$), here in the parameter space $d$ and $a_1$. The shaded areas indicate the regions where the ion tubes are stable. The value of tube radius depends on both $d$ and $a_1$ through Eq.~(\ref{eq:trap-cond}). The inset displays a zoom where the parameters of the plots in Fig. \ref{fig:3d-spect} are shown. \label{fig:stab6}}
\end{figure}

The analysis of the eigenvalues of matrix $D$ as a function of the axial distance $d$ and of the trap parameter $a_1$ allows us to determine a stability diagram for the tubular structures. We find the stability regions by identifying the parameters for which the eigenvalues of $D$ are positive. The borders correspond to the case where one (or more) eigenvalue vanishes, while outside of the stability region there are negative eigenvalues, i.e., the tube is unstable. The stability diagrams were calculated for up to 20 ions per ring for different multipolar orders~$k$. Figure~\ref{fig:stab6} displays one such diagram for the case of an octupole trap. Depending on $d$ and $a_1$ there are different numbers of distinct regions where the tubes are stable. The corresponding stability boundaries are linear when plotted as a function of the radius $R$ and of the distance $d$. This behaviour can be understood in terms of scaling arguments: in fact, the equilibrium configuration can be rescaled by a factor $\lambda$ if the trapping parameter $a_1$ is rescaled by a factor $\lambda^{-(2k-1)}$. In general, for smaller~$N_R$ and higher multipole orders~$k$ there are up to two different stability regions, while for large~$N_R$ and small~$k$ one finds one or no stability region. The number of different stability regions is given in Table~\ref{table:stab} as a function of the multipole order $k$.

\begin{table}
\center
\caption{Number of separate stability regions of the tubes for different multipolar orders~$k$ and number of ions per ring~$N_R$, taking $N_R\ge 4$. For low~$N_R$ two stability regions are found. \label{table:stab}}
\begin{tabular}{|r||l|l|l|l|l|l|l|}
\hline
$2k$ & 6 & 8 & 10 & 12 & 16 & 20 & 22 \\
\hline
\hline
2 regions - $N_R$ & 4-7 & 4-12 & 4-15 & 4-16 & 4-16 & 4-16 & 4-16 \\
\hline
1 region - $N_R$ & 8-13\footnote{For $N_R \geq 14$ no stable region was found.} & 13-20 & 16-20 & 17-20 & 17-20 & 17-20 & 17-20 \\
\hline
\end{tabular}
\end{table}

\subsection{Normal modes}

In the rest of this section we discuss the normal modes of the ion tube. For this purpose, we employ a coordinate system that matches the symmetry of the crystal. Thus, we choose for each particle $\alpha$ a coordinate system such that $x_\alpha$ denotes the coordinate in the radial direction and $y_\alpha$ the coordinate in the azimuthal direction. The potential energy in the new coordinates reads~\cite{Ashcroft}
\begin{align}
 V' = E_0 + \frac{1}{2} \sum_{\alpha, \alpha'} \tilde {\bm u}_\alpha \tilde{D} (\alpha,\alpha') \tilde {\bm u} _{\alpha'}
\end{align}
where
\begin{align}
 \tilde{\bm u}_\alpha = Q^{-1}(\alpha) \, \tilde{\bm u}_\alpha
\end{align}
are the coordinates in the new rotated system and
\begin{align}
\tilde{D} (\alpha,\alpha') = Q^{-1}(\alpha) \, D (\alpha,\alpha') \, Q(\alpha')
\end{align}
are the new matrix coefficients. The rotation matrix used here is given by
\begin{align}
 Q(j,\sigma) = \begin{pmatrix}
  \cos(\phi_{j\sigma}) & -\sin(\phi_{j\sigma})  & 0 \\
  \sin(\phi_{j\sigma}) & \phantom{-}\cos(\phi_{j\sigma}) & 0 \\
  0 & 0 & 1
 \end{pmatrix} ,
\end{align}
where $\phi_{j\sigma} = \phi_0 ( j + \sigma/2 )$. Clearly, the transformation does not depend on the index $l$ labeling the pair cell along the $z$ axis. 
The equations of motion for each ion read
\begin{align} \label{eq:eq-motion}
 m \ddot{\tilde{\bm u}}_\alpha = - \sum_{\alpha'} \tilde{D} (\alpha,\alpha') \tilde{\bm u}_{\alpha'}.
\end{align}
Because of the periodic boundary conditions, this equation can be solved using the ansatz
\begin{align} \label{eq:elong}
 \tilde{\bm u}_{\alpha}^{(k_1,k_2,n)} = {\bm \epsilon}_{\sigma} \exp \left[ {\rm i} \left(\frac{2 \pi k_1 j}{N_R} + \frac{2 \pi k_2 l}{N_A} - \omega_{k_1,k_2,n} t \right) \right] ,
\end{align}
where $k_1$ and $k_2$ take the discrete values $k_1=0, \ldots, N_R-1$ and $k_2=0, \ldots, N_A-1$. Plugging Eq.~(\ref{eq:elong}) into Eq.~(\ref{eq:eq-motion}) leads to an eigenvalue problem, such that for every pair of values $(k_1,k_2)$ one finds the frequencies~$\omega_{k_1,k_2,n}$, with $n=1,\ldots,6$, whose corresponding eigenvectors are $\bm \epsilon_\sigma$. 

\begin{figure*}[hbt]
  \subfloat[$k_1=0$]{\includegraphics[width=0.25\textwidth]{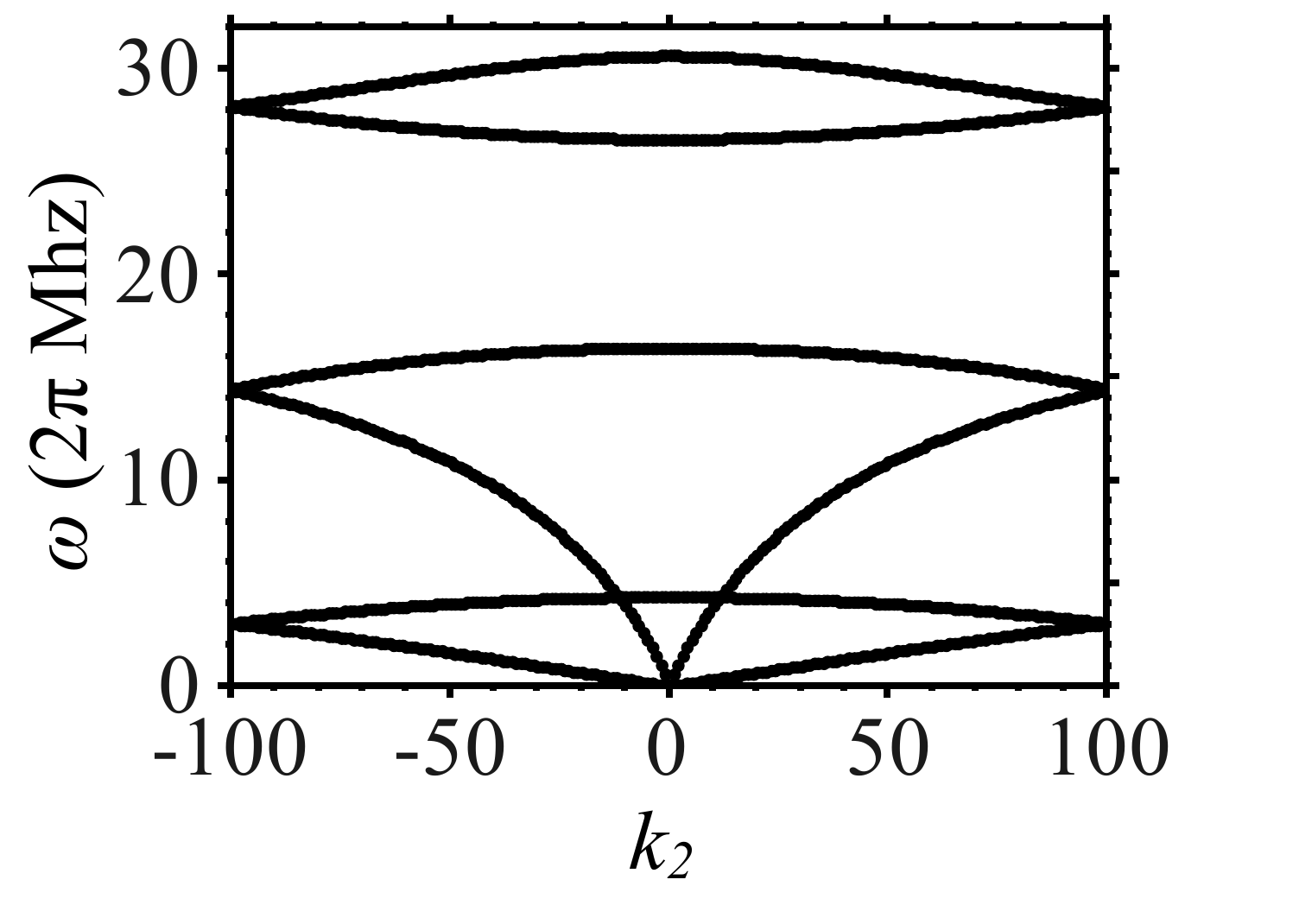}} 
  \subfloat[$k_1=\pm 1$]{\includegraphics[width=0.25\textwidth]{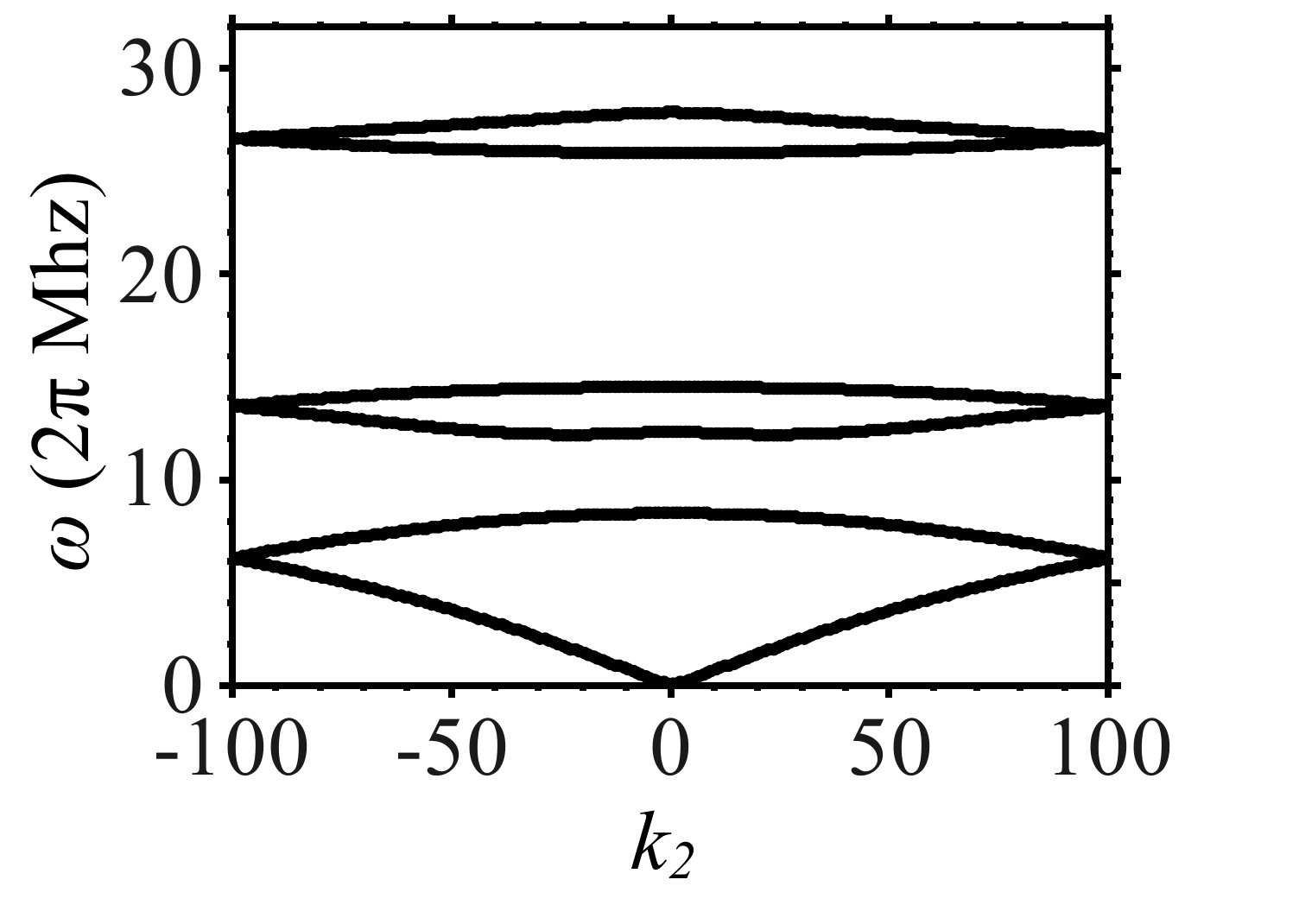}} 
  \subfloat[$k_1=\pm 2$]{\includegraphics[width=0.25\textwidth]{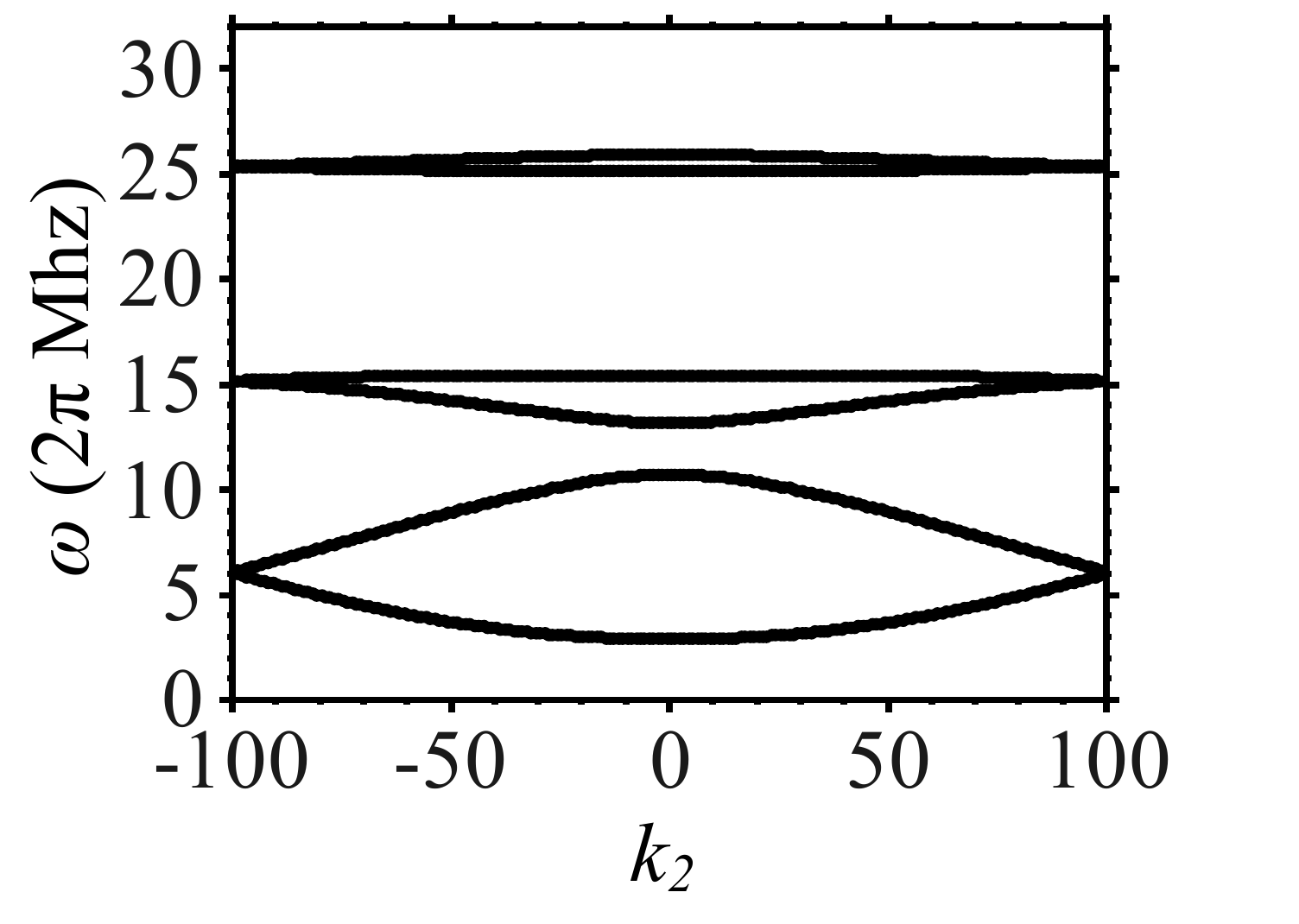}} 
  \subfloat[$k_1=\pm 3$]{\includegraphics[width=0.25\textwidth]{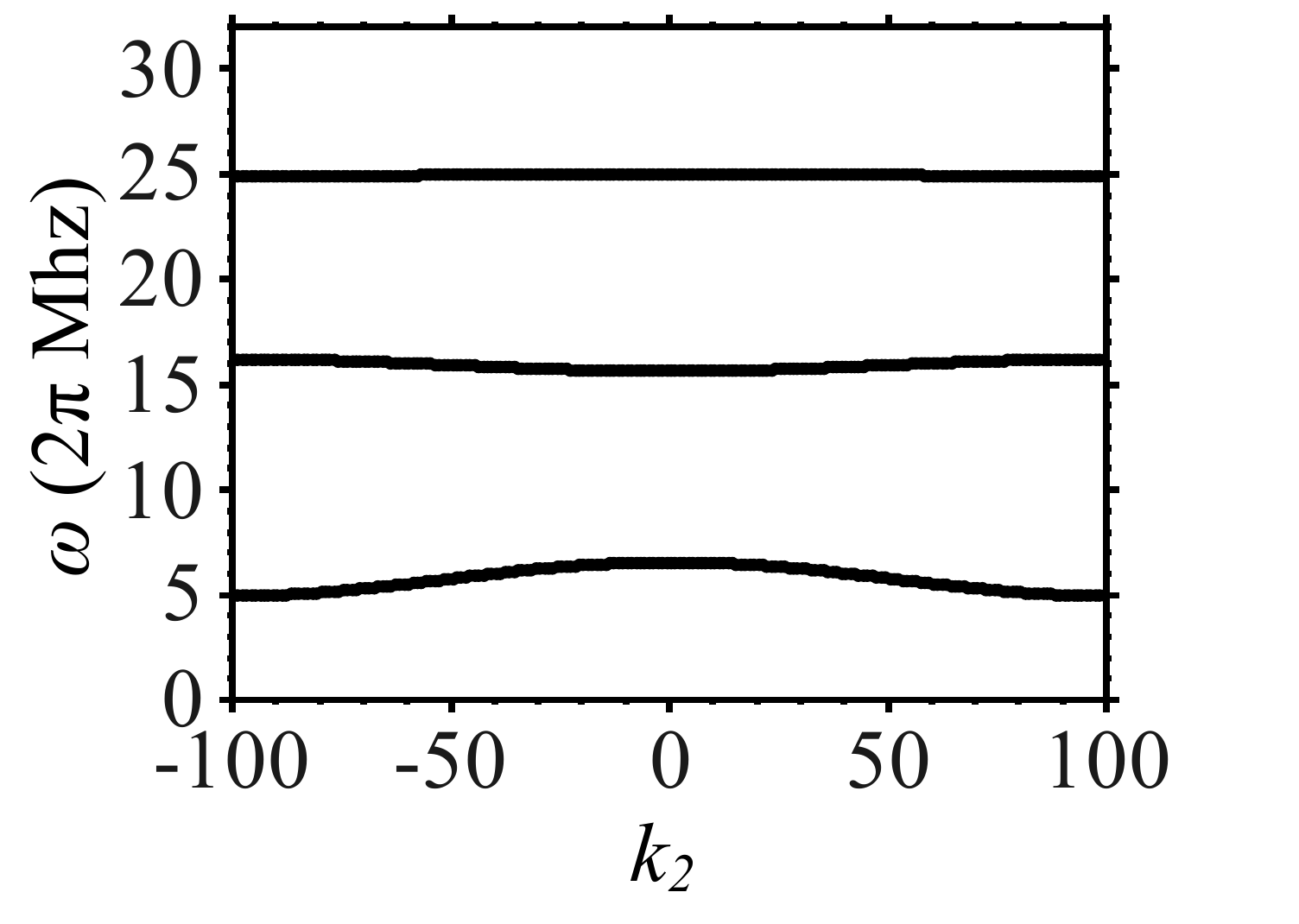}}  \\
  \subfloat[$k_1=0$]{\includegraphics[width=0.25\textwidth]{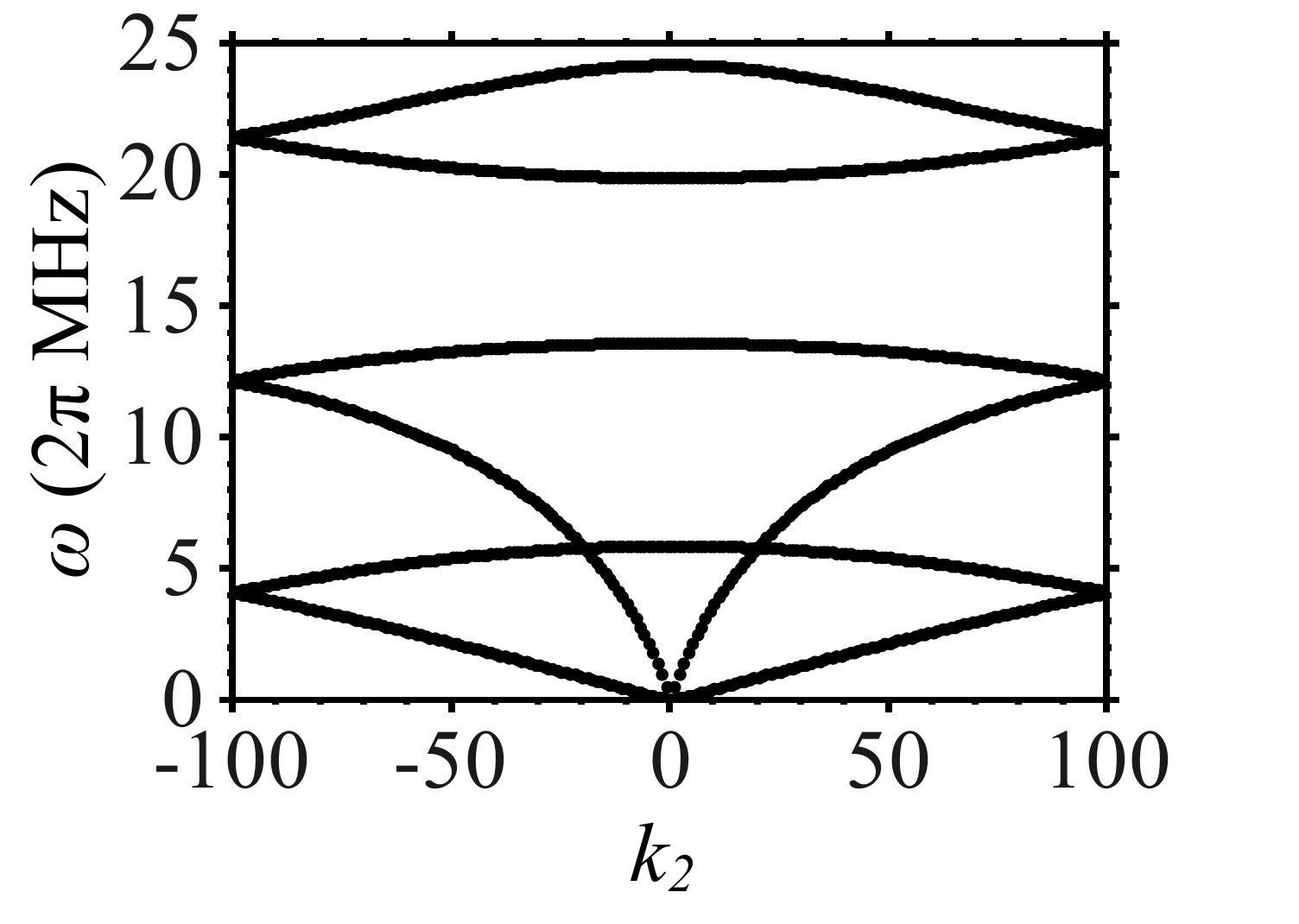}} 
  \subfloat[$k_1=\pm 1$]{\includegraphics[width=0.25\textwidth]{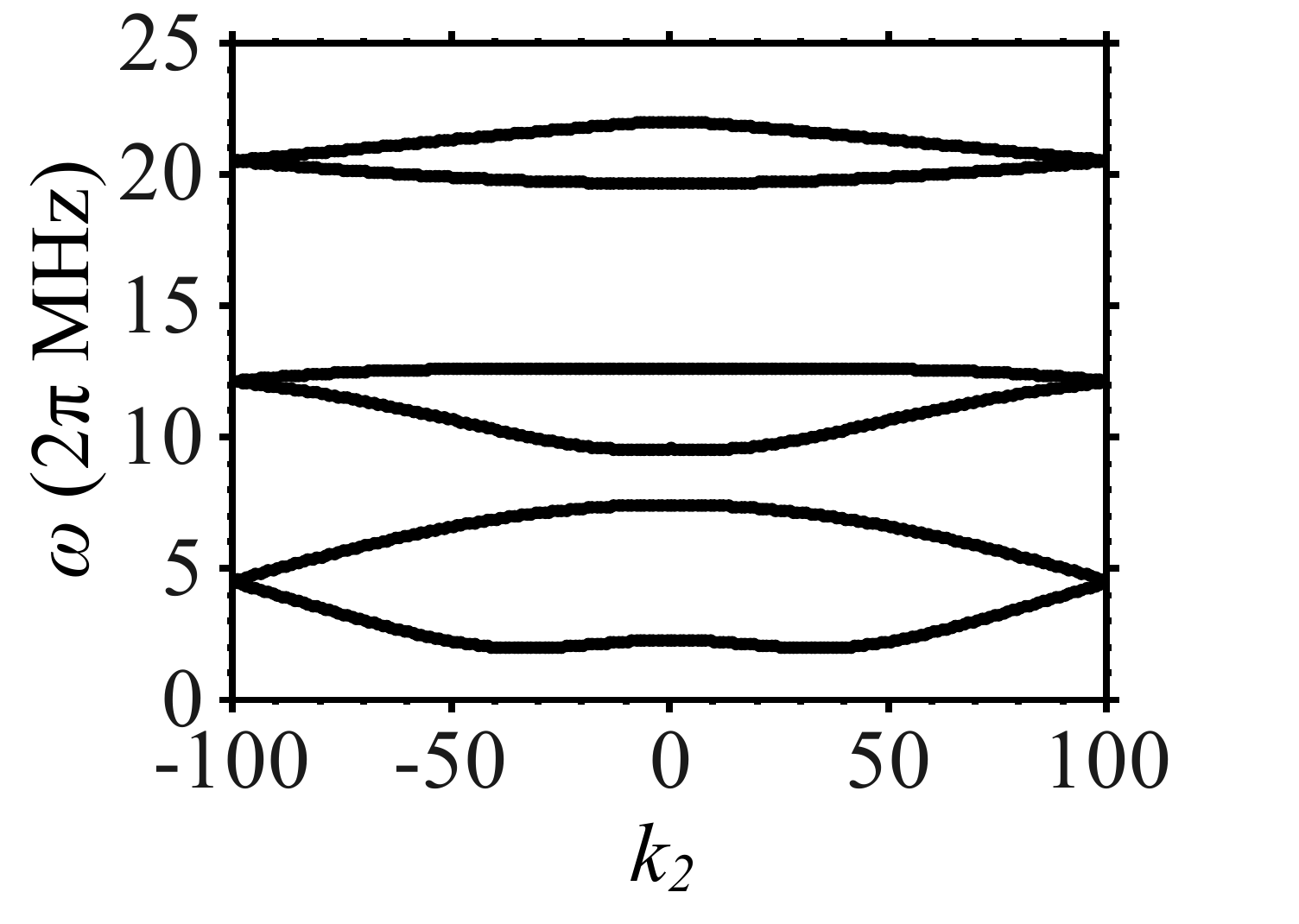}} 
  \subfloat[$k_1=\pm 2$]{\includegraphics[width=0.25\textwidth]{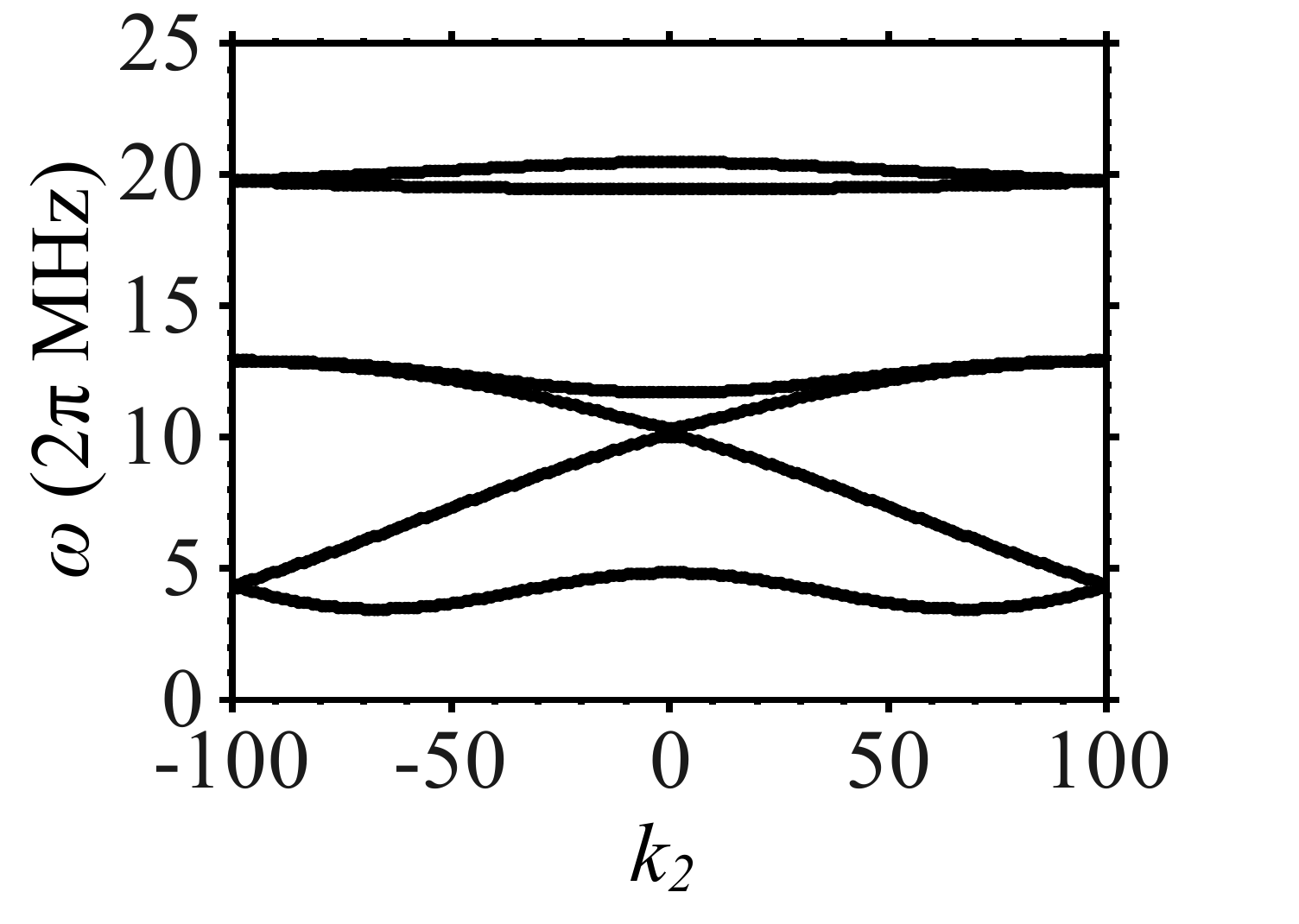}} 
  \subfloat[$k_1=\pm 3$]{\includegraphics[width=0.25\textwidth]{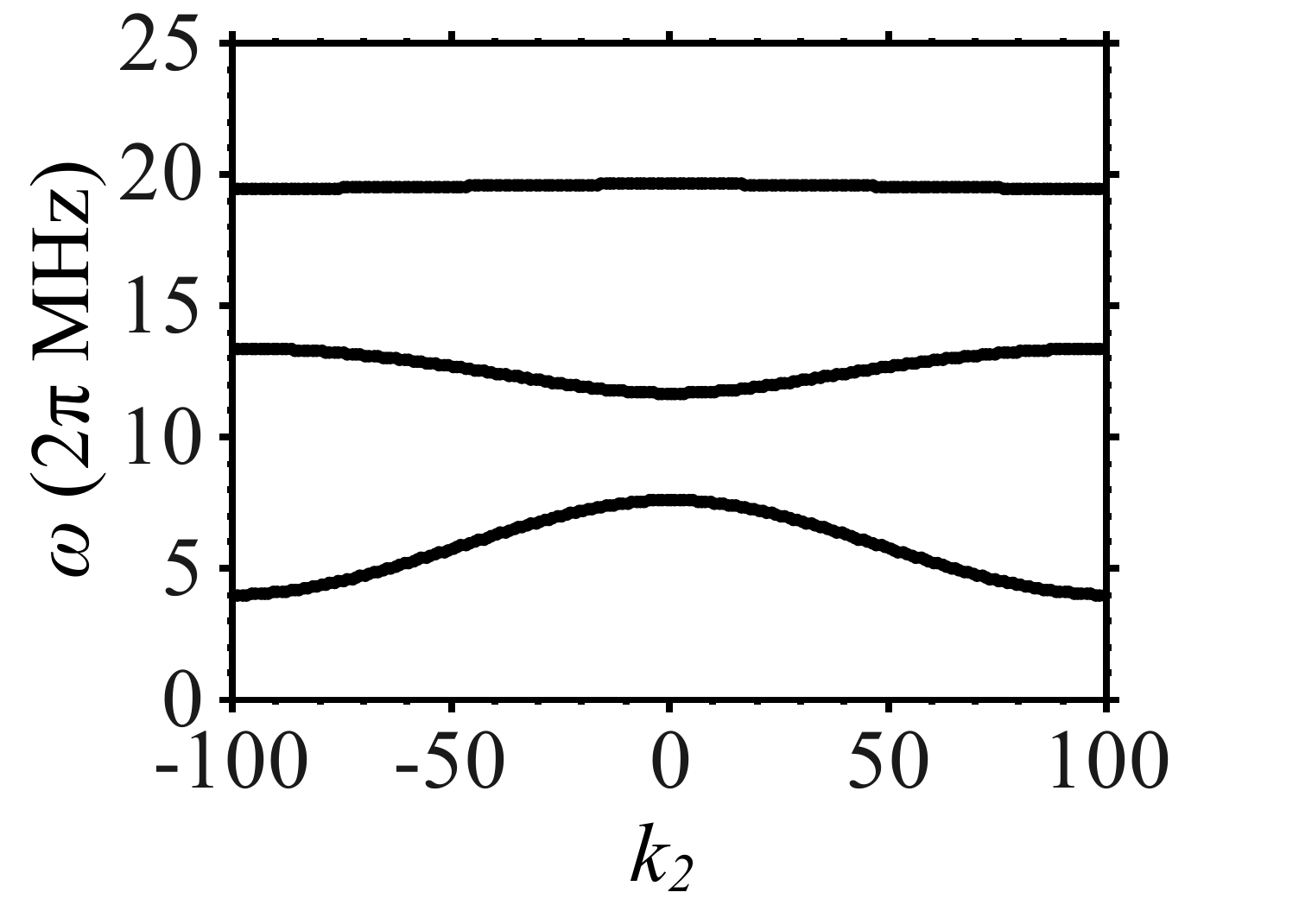}} \\
  \subfloat[$k_1=0$]{\includegraphics[width=0.25\textwidth]{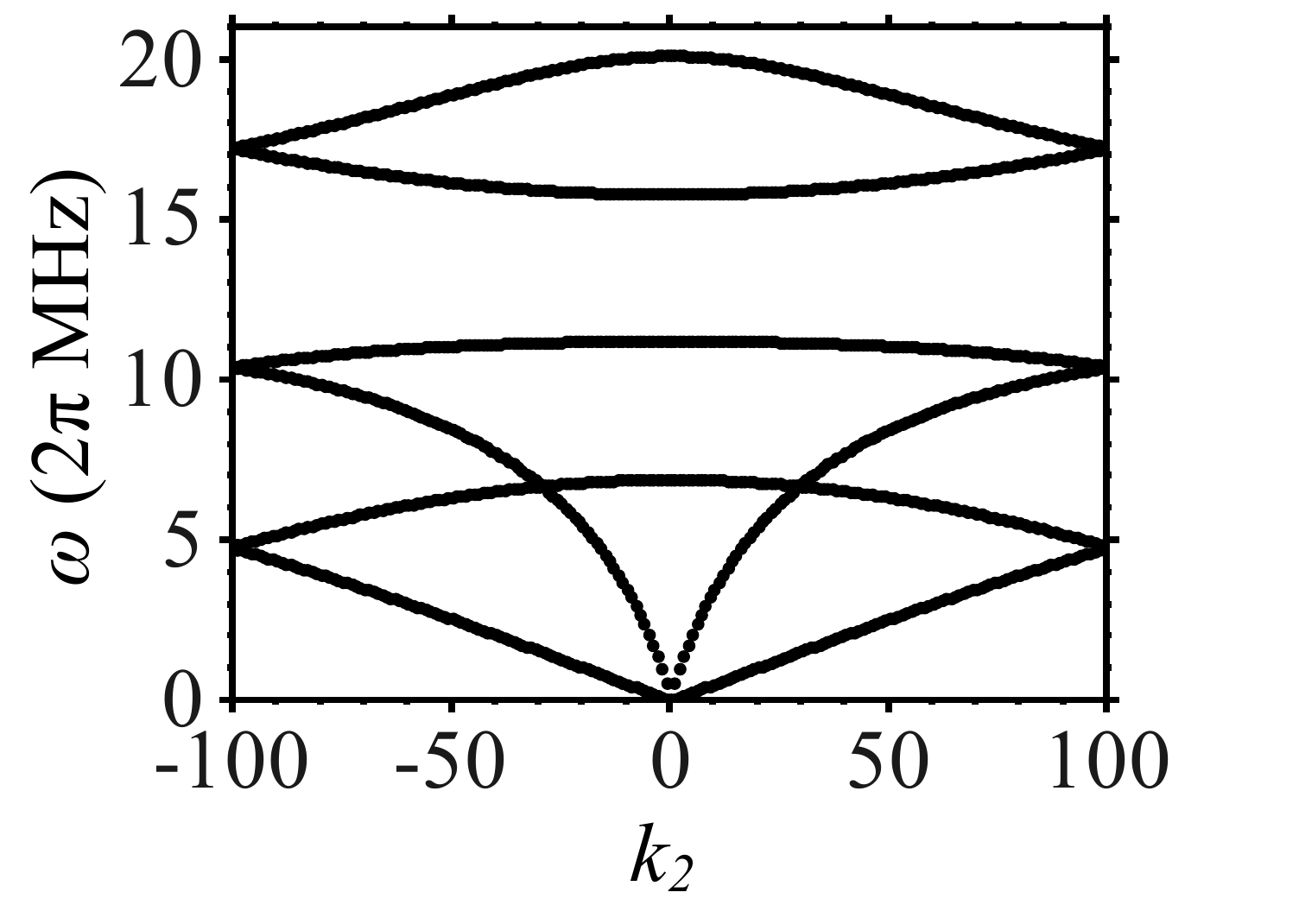}} 
  \subfloat[$k_1=\pm 1$]{\includegraphics[width=0.25\textwidth]{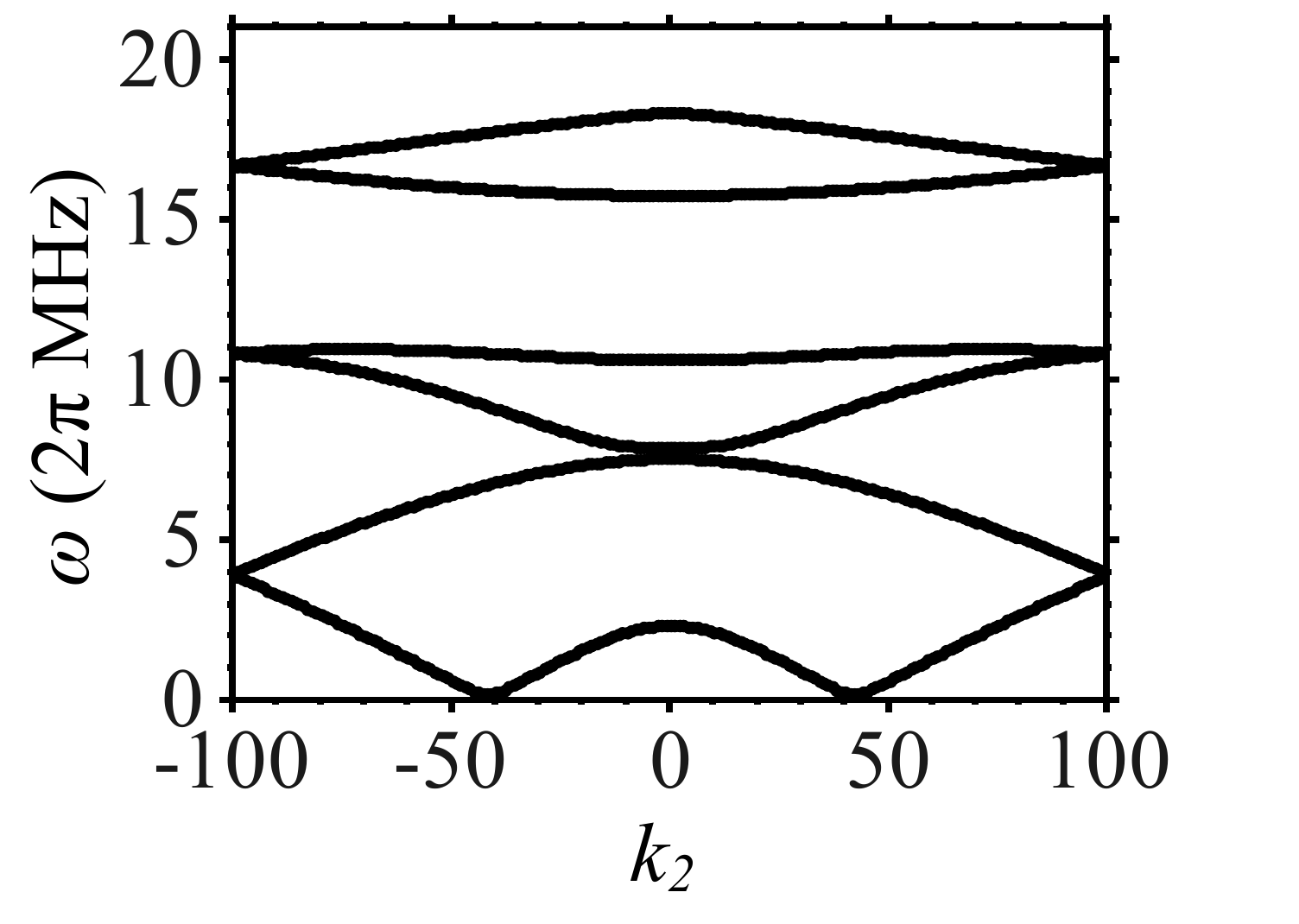}} 
  \subfloat[$k_1=\pm 2$]{\includegraphics[width=0.25\textwidth]{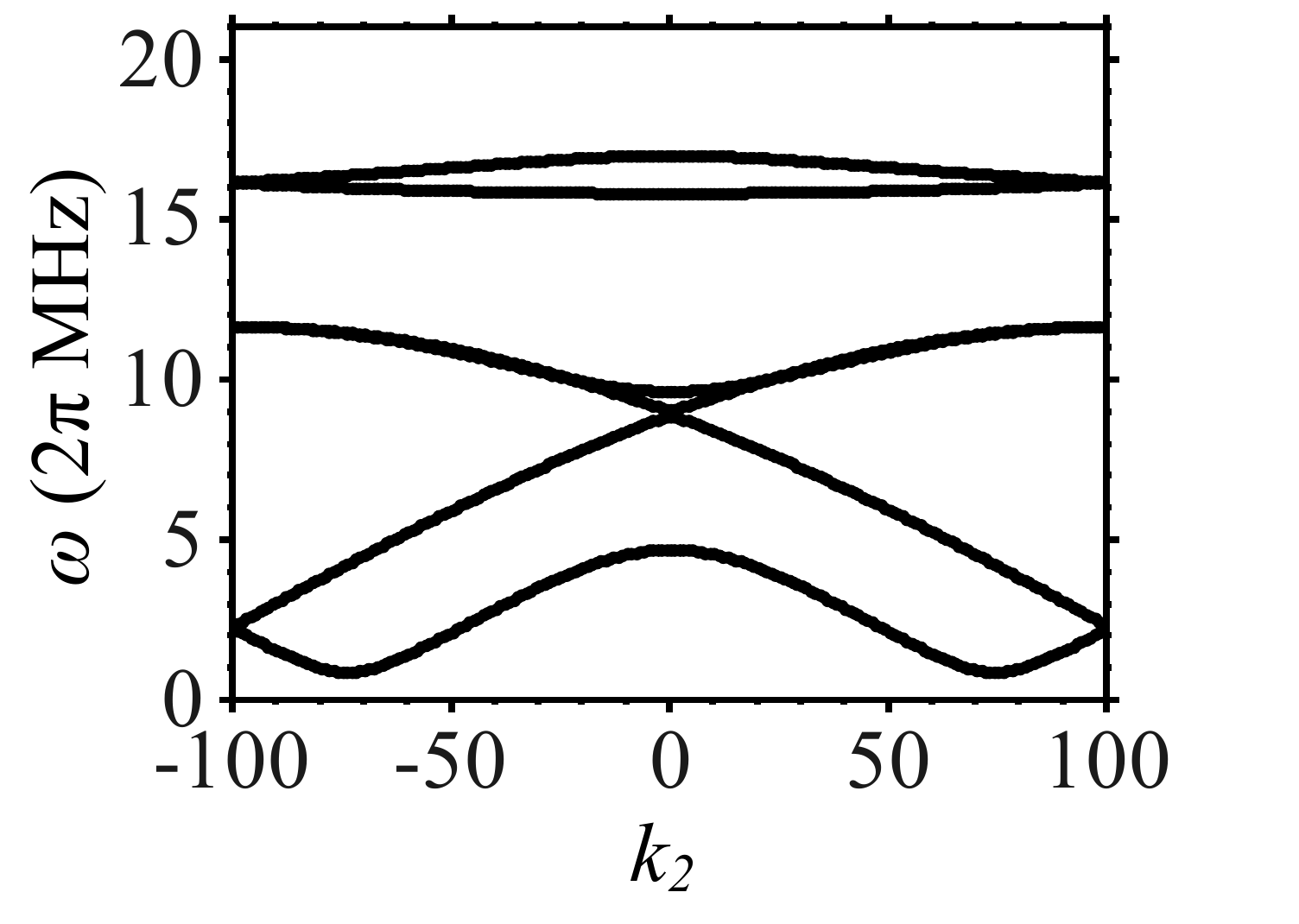}} 
  \subfloat[$k_1=\pm 3$]{\includegraphics[width=0.25\textwidth]{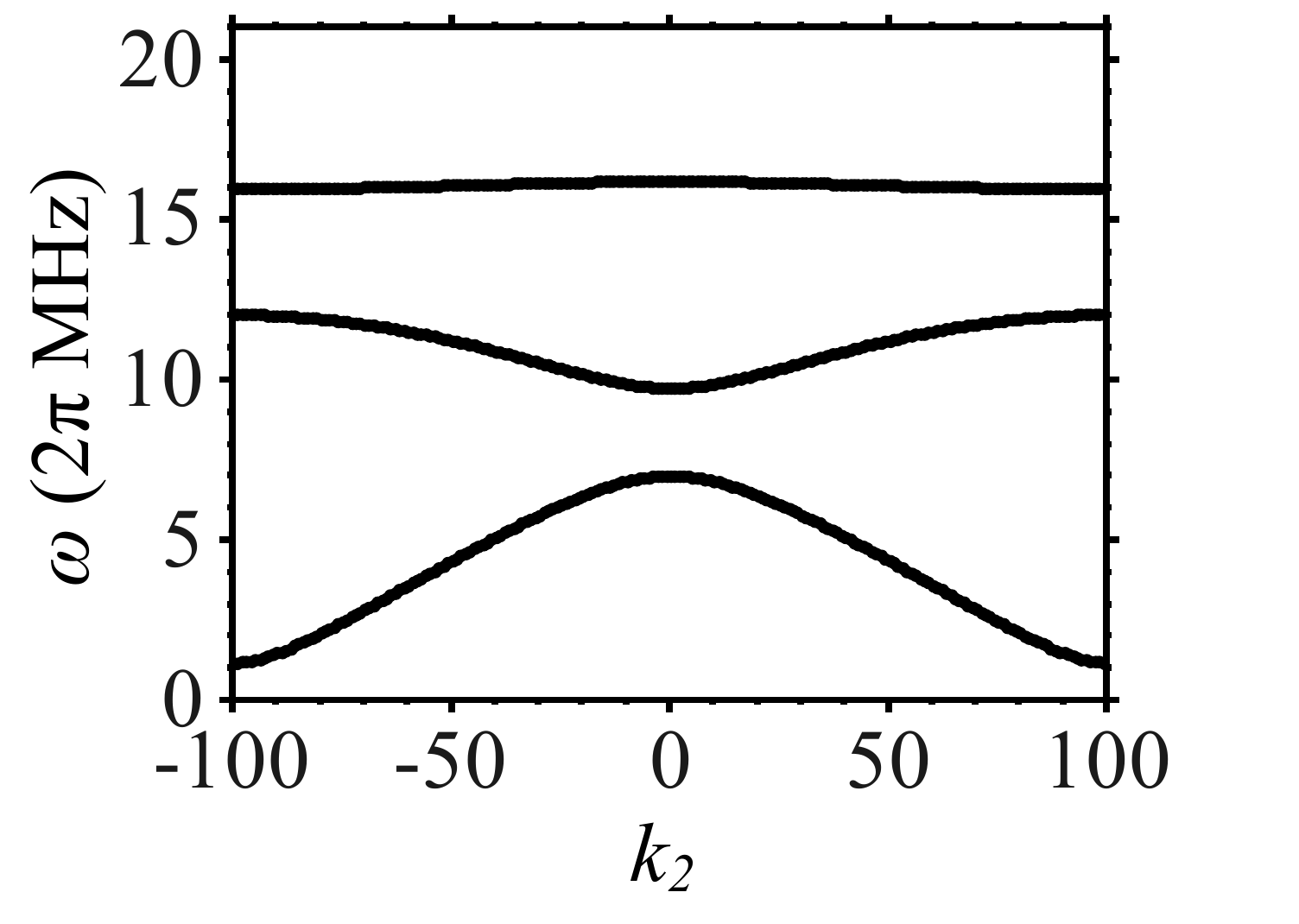}}
 \caption{\label{fig:3d-spect} Spectra of 400 rings of Ca$^{+}$ ions in an octupole trap for different trap parameters. Each ring is composed by six ions, the distance between rings is $10\mu$m. The first row,~(a)-(d), shows the spectrum for $a_1 = 3.4\cdot 10^{9}$ J m$^{-6}$ (upper point in the inset of Fig. \ref{fig:stab6}), the second row,~(e)-(h) for $a_1 = 8.4\cdot 10^{8}$ J m$^{-6}$ (central point in Fig. \ref{fig:stab6}), and the third row~(i)-(l) for $a_1 = 2.7\cdot 10^{8}$ J m$^{-6}$ (lower point in Fig. \ref{fig:stab6}).  Each column corresponds to a value of $k_1$: from left to right $k_1=0,\pm 1, \pm 2, \pm 3$. Because of the specific symmetry of the crystal, for $k_1=N_R/2$ (when $N_R$ is even) each frequency is twofold degenerate at a fixed value of $k_2$.}
\end{figure*}

The eigenfrequencies were calculated numerically for different values of $N_A$, $N_R$, $d$, and $R$. Typical spectra for a fixed distance $d=10\mu$m are shown in Fig.~\ref{fig:3d-spect}. The spectra are evaluated for $400$ rings, but we note that, by comparing the spectra for different values of the number of rings $N_A$, one finds that tubes composed by tens of rings have a spectrum which is already close to the one reported here. 

Each row in Fig.~\ref{fig:3d-spect} corresponds to a different steepness of the radial trapping potential, $a_1$, while each column corresponds to the spectra obtained for values of $k_1=0,\pm 1,\pm 2,\pm 3$. In particular $a_1$ decreases from top to bottom: the value in the first row (from above) corresponds to the parameters of the upper point in the inset of Fig. \ref{fig:stab6}, the second row corresponds to the parameters of the central point, and the third row to the parameters of the point at the lower edge of the stability region.  

We observe the existence of 6 branches for each value of $k_1$. They are reduced to three (each twofold degenerate) for $k_1=\pm 3$. Because of the continuous rotational and translational symmetries of the trapping potential, two eigenfrequencies at $k_1 = k_2 = 0$ are zero: one of these corresponds to the rigid rotation about the $z$-axis of the crystal, the other to rigid translations along $z$. In general, the eigenmodes at $k_1=0$ can be understood in terms of normal modes of a two-dimensional biperiodic crystalline structure~\cite{Ashcroft}. For $k_2=0$, on the other hand, the eigenmodes are independent of the axial coordinate $z$. We denote the eigenmodes with $k_1=k_2=0$ by bulk excitations.

The tube becomes unstable at the points where eigenfrequencies, other than the bulk eigenfrequencies, vanish. For the cases we studied the corresponding modes do not offer an intuitive interpretation. For some multipole orders and numbers of ions per ring, they correspond to motion in azimuthal and axial direction suggesting a reordering of the structure within a cylinder without changing the radius. 

\section{Conclusions} \label{sec:conclusions}

The stability of ion rings in linear multipole traps was studied both analytically and numerically. Two types of clusters were considered: single rings, which can form in the center of the linear multipole trap, and ion tubes, which are found for large numbers of ions when the axial potential is sufficiently flat. For the latter case, the stability diagram and the normal mode spectrum in the stability region were determined.  

This study sets the basis for a thermodynamic characterization of these structures in the spirit of the work in Ref. \cite{Morigi2004}. Such study will require the definition of an appropriate thermodynamic limit, which can be identified with the number of rings going to infinity, $N_A \to \infty$, while the distance~$d$ between rings and the radius~$R$ remain constant, meaning that the linear density remains constant. A different thermodynamic limit can be identified for a single ring structure. This will in fact approach a linear chain in the theoretical limit in which the radius of the ring becomes infinite while the ions density on the ring is kept constant. An experimental situation deals with finite size systems, which include the effect of the curvature of the ring. For odd numbers of ions, the instability of the ring leads to the creation of localized defects, which  can be considered for quantum information platform with localized kinks, in the spirit of the proposal in Ref. \cite{Reznik2010}. 

The influence of quantum fluctuations at the instability between a single and a double ring, furthermore, can provide additional aspects to the one identified in Ref. \cite{Shimshoni2011} for the case of a linear chain in a linear quadrupole trap. 

\acknowledgements The authors are grateful to Caroline Champenois and Martina Knoop for stimulating discussions. We acknowledge support by the European Commission (Integrating Project ``AQUTE'', STREP ``PICC'', COST action ``IOTA''), the Alexander von Humboldt and the German Research Foundations.  F.C. and G.M. thank the group ``Confinement d'Ion et Manipulation Laser'' at the Universit\'e de Provence in Marseille for the kind hospitality. G.M. acknowledges support of a guest professorship from the  Universit\'e de Provence in Marseille.

\begin{appendix}

\section{}

\
 \begin{figure*}
 %\subfloat[$\Gamma=0.03,2,10^4$, $2k=10$]{\includegraphics[width=0.3\textwidth]{./density10000k5.pdf}}
%  \subfloat[$\Gamma=2$, $2k=10$]{\includegraphics[width=0.3\textwidth]{./density2k5.pdf}}
 \subfloat[$\gamma=0.029$, $2k=10$]{\includegraphics[width=0.3\textwidth]{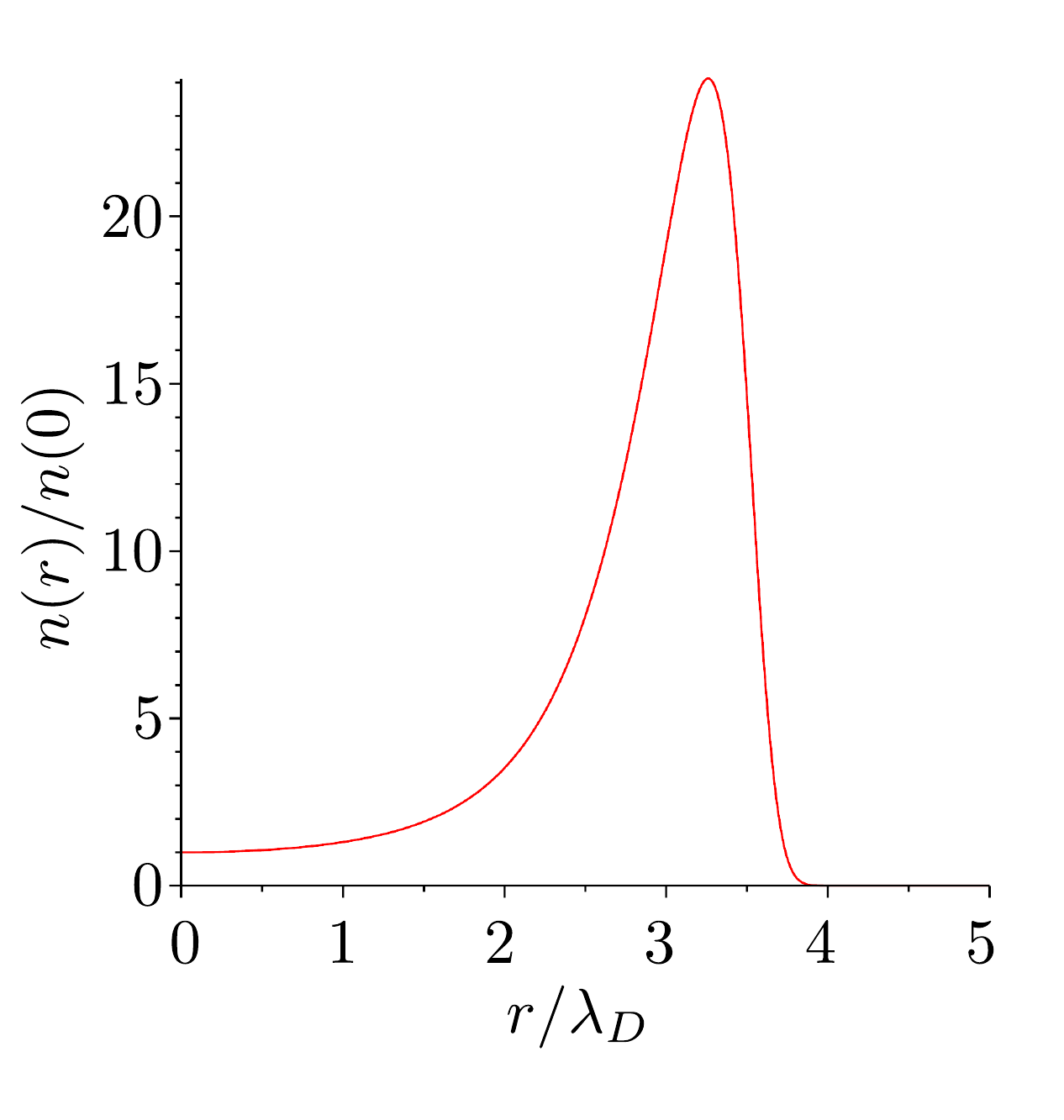}}
 %\subfloat[$\Gamma=1.5,10,100$, $2k=4$]{\includegraphics[width=0.3\textwidth]{./density100k2.pdf}}
 % \subfloat[$\Gamma=10$, $2k=4$]{\includegraphics[width=0.3\textwidth]{./density10k2.pdf}}
 \subfloat[$ \gamma=1.5$, $2k=4$]{\includegraphics[width=0.3\textwidth]{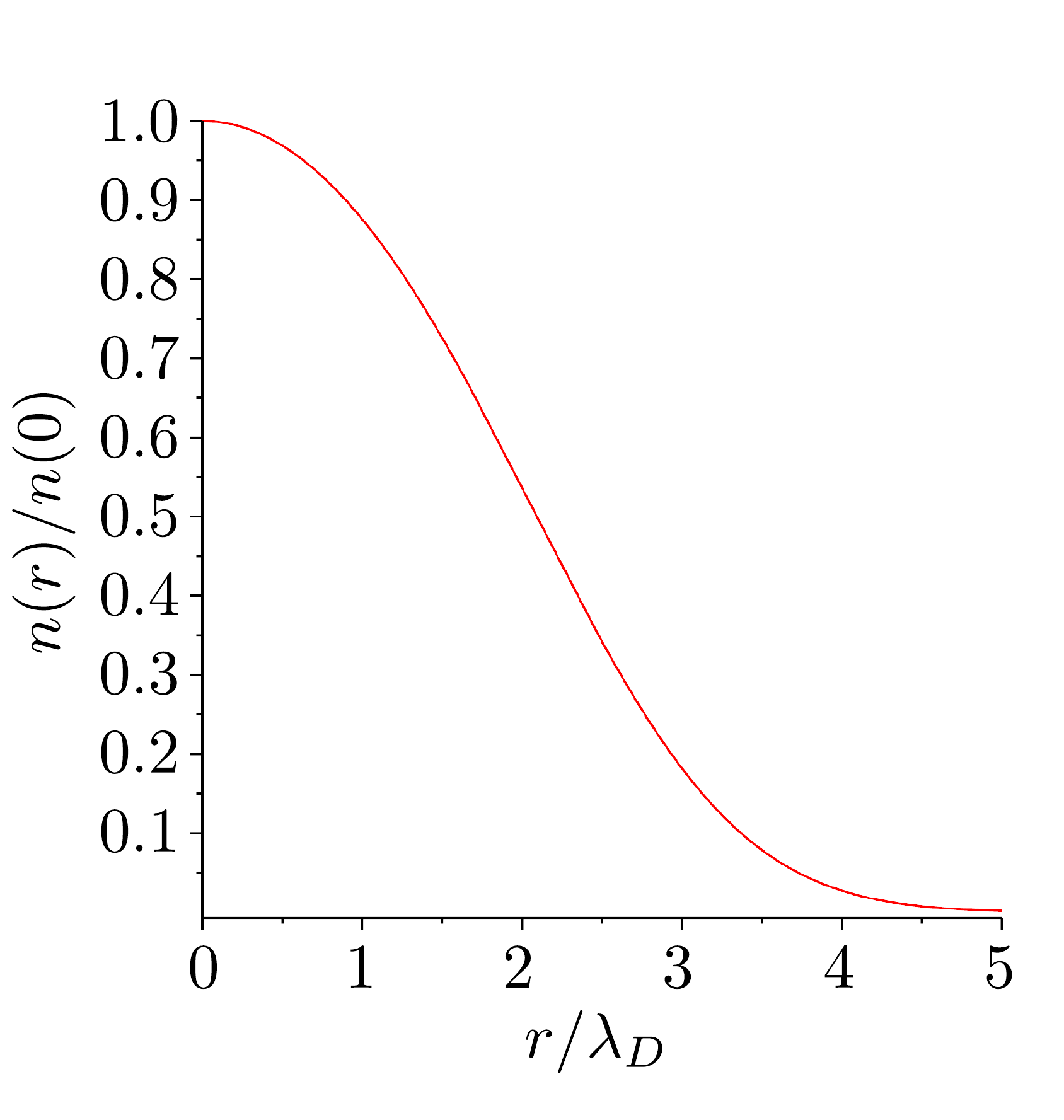}} 
 \caption{Density profiles in (a) a multipole trap with $k=5$ and  (b) a quadrupole trap~($k=2$) for different values of the parameter~$\gamma$.}
  \label{fig:density}
 \end{figure*}
In order to find the density profile, the two coupled equations~\eqref{eq:boltzmann}-\eqref{eq:poisson} need to be solved. This can be done in a straightforward way in the case of an infinitely long trap~\cite{Champenois2009}.  
 It is convenient to introduce the logarithmic density profile
 \begin{align}
  \Psi(\bm r) = \ln \left[ \frac{n(\bm r)}{n(\bm 0)} \right],
 \end{align}
 and express the density as a function of this quantity,
 \begin{align} \label{eq:density-psi}
  n(\bm r) = n(\bm0) \exp\left[ \Psi(\bm r) \right] \,.
 \end{align}
 By inserting eq.~(\ref{eq:density-psi}) in the Poisson eq.~(\ref{eq:poisson}), one finds
 \begin{align} \label{eq:poisson-psi}
  \Delta \Psi (\bm r) = \frac{1}{\lambda_D^{(0)}} \left\{ \exp\left[ \Psi(\bm r) \right] - \frac{\epsilon_0}{q^2 n(\bm 0)} \Delta V_{\rm trap} \right\}
 \end{align}
 where we call $\lambda_D^{(0)}$ the Debye length in the center of the trap. We note that this equation only depends on the effective potential of the trap through the Laplacian; since the static potential $V_{st}$ must satisfy the Laplace equation, it has no influence on the density profile within the region of validity of approximation (\ref{eq:boltzmann}).
 In rescaled cylindrical coordinates $\rho = r/\lambda_D^{(0)}$ and $\xi = z/\lambda_D^{(0)}$, and writing the explicit form of $V_{\rm rf}$, Eq.~(\ref{eq:poisson-psi}) becomes
 \begin{align}
  \left( \frac{1}{\rho} \frac{\partial}{\partial \rho} \rho \frac{\partial}{\partial \rho} + \frac{\partial^2}{\partial \xi ^2} \right) \Psi (\rho, \xi) = \exp\left[ \Psi(\rho,\xi)\right] - \gamma \rho^{2k-4},
 \end{align}
 with the parameter
 \begin{align}
  \gamma = \frac{\epsilon_0}{n_0} \frac{1}{4m} \left[\frac{V k (2k-2)}{\Omega r_0^{k}}\right]^2 {\lambda_D^{(0)}}^{2k-4} .
 \end{align}
For a very long cylinder, the $\xi$ dependence can be neglected and the equation can be integrated. The results of this integration are shown in Fig.~\ref{fig:density}. While for small Debye lengths (small $\gamma$) in a quadrupole trap~($2k=4$) there is a nearly constant density in the center of the trap, in any multipole trap~($2k>4$) the density increases from the center, reaches a maximum and then drops to nearly zero within one Debye length. Large Debye lengths, corresponding to relatively high temperatures of the plasma, result in density profiles in multipole traps which are nearly constant over the whole charge distribution.
\end{appendix}


%\bibliography{papers}

\begin{thebibliography}{99}
\bibitem{footnote}In quadrupole traps, for $N>5$ the single-ring configuration is not a global minimum, but can be a local minimum.
\end{thebibliography}

\end{document}